\documentclass[a4paper]{packages/paper_cw} 
\usepackage{helvet}                 

\usepackage{amsmath}                
\usepackage{amssymb}                
\usepackage{mathtools}              
\usepackage{bm}                     

\usepackage{graphicx}               
\usepackage{float}                  
\usepackage{booktabs}               
\usepackage{multirow}               
\usepackage[table,xcdraw]{xcolor}   
\usepackage{colortbl}               
\usepackage{subfig}                 
\usepackage{placeins}               
\usepackage[page]{appendix}         

\usepackage{tikz}                   
\usetikzlibrary{arrows.meta,positioning,shapes.geometric,fit}
\usepackage{pgfplots}               
\pgfplotsset{compat=1.18}           

\usepackage[numbers]{natbib}        

\usepackage{siunitx}                

\usepackage{enumitem}               
\usepackage{comment}                
\usepackage{listings}               
\usepackage{lineno}                 

\usepackage{nomencl}                

\usepackage{hyperref}               
\hypersetup{
  colorlinks=true,
  linkcolor=red,
  citecolor=blue,
  urlcolor=blue,
  pdfauthor={Vahab Narouie},
  pdftitle={Your Title}
}

\usepackage{amsthm}

\makenomenclature
\newcommand{\Rev}{\textcolor{black}} 

\long\def\symbolfootnote[#1]#2{\begingroup \def\thefootnote{\fnsymbol{footnote}}\footnote[#1]{#2} \endgroup}

\newcommand{\scl}[1]{ \ensuremath{  #1  } }
\renewcommand{\vec}[1]{ \ensuremath { \mathbf{ #1 } } }
\newcommand{\gvec}[1]{ \ensuremath { \boldsymbol{ #1 } } }
\newcommand{\mat}[1]{ \ensuremath { \pmb{\mathbf{ #1 } } } }
\newcommand{\ten}[1]{ \ensuremath {  \pmb{\mathbf{ #1 } } }}

\newcommand{\scln}{\ensuremath{ \scl{n} }}

\newcommand{\vecb}{\ensuremath{ \vec{b} }}

\newcommand{\vece}{\ensuremath{ \vec{e} }}

\newcommand{\vecp}{\ensuremath{ \vec{p} }}

\newcommand{\vecr}{\ensuremath{ \vec{r} }}

\newcommand{\vect}{\ensuremath{ \vec{t} }}
\newcommand{\vecu}{\ensuremath{ \vec{u} }}
\newcommand{\vecv}{\ensuremath{ \vec{v} }}

\newcommand{\vecy}{\ensuremath{ \vec{y} }}
\newcommand{\vecz}{\ensuremath{ \vec{z} }}

\newcommand{\vecN}{\ensuremath{ \vec{N} }}

\newcommand{\vecX}{\ensuremath{ \vec{X} }}

\newcommand{\veckappa     }{\ensuremath{ \gvec{\kappa} }}

\newcommand{\vecmu        }{\ensuremath{ \gvec{\mu} }}

\newcommand{\vecsigma     }{\ensuremath{ \gvec{\sigma} }}

\newcommand{\vecvarepsilon}{\ensuremath{ \gvec{\varepsilon} }}

\newcommand{\matA}{\ensuremath{ \mat{A} }}

\newcommand{\matC}{\ensuremath{ \mat{C} }}

\newcommand{\matF}{\ensuremath{ \mat{F} }}

\newcommand{\matH}{\ensuremath{ \mat{H} }}
\newcommand{\matI}{\ensuremath{ \mat{I} }}

\newcommand{\bbE}{\ensuremath{ \mathbb{E} }}

\newcommand{\tenE}{\ensuremath{ \ten{E} }}
\newcommand{\tenF}{\ensuremath{ \ten{F} }}

\newcommand{\tenH}{\ensuremath{ \ten{H} }}
\newcommand{\tenI}{\ensuremath{ \ten{I} }}

\newcommand{\tenP}{\ensuremath{ \ten{P} }}

\newcommand{\tenS}{\ensuremath{ \ten{S} }}

\newcommand{\tent}{\ensuremath{ \ten{t} }}

\renewcommand{\ln}[1]{\text{$\hspace{0.1cm}$ln$\left(#1\right)$}}

\DeclareMathOperator*{\argmin}{arg\,min}

\newcommand{\norm}[1]{\left\lVert#1\right\rVert}

\renewcommand{\vec}[1]{ \ensuremath{ \mathbf{#1} } }

\renewcommand{\ln}[1]{\text{ln$\left(#1\right)$}}

\makeatletter
\renewcommand{\fnum@figure}{\textbf{Fig. \thefigure}}
\renewcommand{\fnum@table}{\textbf{Tab. \thetable}}
\makeatother

\newcommand{\physicalDomain}{\ensuremath{\Omega}}

\NewDocumentCommand{\expectOper}{ O{} }{%
	\ifx&#1&%
	\ensuremath{\bbE}%
	\else
	\ensuremath{\bbE \left[#1\right]}%
	\fi
}

\NewDocumentCommand{\meanFunc}{ O{} m }{%
	\ifx&#1&%
	\ensuremath{\mu_{#2}}%
	\else
	\ensuremath{\mu_{#1}(#2)}%
	\fi
}

\NewDocumentCommand{\covFunc}{ O{} m }{%
	\ifx&#1&%
	\ensuremath{C_{#2}}%
	\else
	\ensuremath{C_{#1}(#2,#2')}%
	\fi
}

\NewDocumentCommand{\corFunc}{ O{} m }{%
	\ifx&#1&%
	\ensuremath{R_{#2}}%
	\else
	\ensuremath{R_{#1}(#2,#2')}%
	\fi
}

\NewDocumentCommand{\stdFunc}{ O{} m }{%
	\ifx&#1&%
	\ensuremath{\sigma_{#2}}%
	\else
	\ensuremath{\sigma_{#1}(#2)}%
	\fi
}

\newcommand{\firstPK}{\tenP}
\newcommand{\secondPK}{\tenS}
\newcommand{\stochDisp}{\ensuremath{\vecu}}
\newcommand{\preStochDisp}{\overline{\stochDisp}}

\newcommand{\preDispOnBound}{\partial\physicalDomain_{\stochDisp}}
\newcommand{\preTracOnBound}{\partial\physicalDomain_{\firstPK}}

\newcommand{\unitNormal}{\vecN}

\NewDocumentCommand{\eigenFunction}{ O{} m }{%
	\ifx&#1&%
	\ensuremath{\phi_{#2}}%
	\else
	\ensuremath{\phi_{#1}(#2)}%
	\fi
}

\newcommand{\veczero}{\ensuremath{ \vec{0} }}

\newcommand{\nsen}{\scln_{\text{sen}}}
\newcommand{\ngdof}{\scln_{\text{gdof}}}                            
\newcommand{\dof}{\scln_{\text{dof}}}                            
\newcommand{\ngsen}{\scln_{\text{gsen}}}                            

\newcommand{\PDF}[2]{\ensuremath{f_{#1} (#2)}}

\newcommand{\defGrad}{\ensuremath{\matF}}

\begin{document}
\title{Unsupervised Constitutive Model Discovery from Sparse and Noisy Data}
\author{Vahab Knauf Narouie$^{\star,1}$,  Jorge-Humberto Urrea-Quintero$^{1}$, Fehmi Cirak$^{2}$,  Henning Wessels$^{1}$}
\institute{(1) Institute of Applied Mechanics, Division Data-driven Modeling of Mechanical Systems, Technische Universität Braunschweig, Pockelsstr. 3, 38106 Braunschweig, Germany \\
	\noindent(2) Department of Engineering, University of Cambridge, Trumpington Street, Cambridge CB2 1PZ, United Kingdom \\
	$\star$ \email{\href{mailto:v.narouie@tu-braunschweig.de}{v.narouie@tu-braunschweig.de}} \\
}
\maketitle


\thispagestyle{empty}


\abstract
{
	Recently, unsupervised constitutive model discovery has gained attention through frameworks based on the {Virtual Fields Method} (VFM), most prominently the EUCLID approach. However, the performance of VFM-based approaches, including EUCLID, is affected by measurement noise and data sparsity, which are unavoidable in practice. The {statistical finite element method} (statFEM) offers a complementary perspective by providing a Bayesian framework for assimilating noisy and sparse measurements to reconstruct the full-field displacement response, together with quantified uncertainty. While statFEM recovers displacement fields under uncertainty, it does not strictly enforce consistency with constitutive relations. In this work, we integrate statFEM with unsupervised constitutive model discovery in the EUCLID framework, yielding {statFEM--EUCLID}. The framework is demonstrated for isotropic hyperelastic materials. The results show that this integration reduces sensitivity to noise and data sparsity, while ensuring that the reconstructed fields remain consistent with both equilibrium and constitutive laws.
}

\section{Introduction}

The vast number of constitutive models for solids proposed in the literature immediately raises a central question: how can one select the most appropriate model for a given dataset? Traditionally, this question of model selection is resolved through expert knowledge and iterative trial and error, a process that can be very time-consuming. This is because constitutive modeling is accompanied by the problem of parameter identification: once a constitutive law is postulated, its parameters must be calibrated against the experimental data. The best model is then typically the model that best explains the given data. Constitutive model discovery aims to automate this process by jointly determining both the model structure and its parameters directly from data. Recent approaches range from sparsity-promoting regression methods \cite{flaschel2021unsupervised,flaschel2023automatedbrain,urre2025MDisc}, to symbolic regression \cite{kissas2024language}, to neural-network-based formulations; see \cite{fuhg2024review,romer2025reduced} for comprehensive reviews.

Modern full-field experimental techniques such as {Digital Image Correlation} (DIC) provide displacement and strain fields with high spatial resolution, even under complex loading scenarios \cite{sutton2008DIC,pierron2023MatTesting2,he2024review,bogusz2024application}. However, stresses are, by their nature, not directly observable and are impossible to infer, especially under heterogeneous or non-proportional loadings. This precludes the direct use of supervised model discovery, which requires paired stress--strain data \cite{tu2020stress}, and motivated the development of {unsupervised} discovery approaches \cite{flaschel2021unsupervised}. Unsupervised discovery constitutes an inverse problem, which is composed of a state equation (given by the boundary value problem) and an observation equation that measures the distance between simulations and data.

Finite Element Model Updating (FEMU) \cite{chen2024finite, akerson2025learning} casts the identification problem as a PDE-constrained optimization task, where material parameters are iteratively updated to minimize the discrepancy between simulated and measured displacement fields. In practice, FEMU is typically implemented as a Nonlinear Least-Squares FEM approach, where the mismatch between simulation and measurement is minimized through iterative solvers. While this formulation is highly flexible and can accommodate arbitrary constitutive forms, it requires costly repeated forward finite element solves. \Rev{The most efficient implementation is via adjoints, where, independently of the number of material parameters, only one forward solve per iteration is required. When numerical differentiation via finite differences is used to approximate the derivative of the loss with respect to material parameters, the number of necessary forward calls depends on the order of the finite difference stencil and the number of material parameters. Automatic differentiation can be used to reduce the cost, see e.g. \cite{seidl2022calibration}. }



The {Virtual Fields Method} (VFM) \cite{pierron2012virtual} provides a variational framework that does not require any forward solves. Here, the nodal degrees of freedom are equated with the observational data, and parameters are identified such that the state equation is satisfied within a norm. Building on this principle, the unsupervised constitutive model discovery framework known as EUCLID \cite{flaschel2021unsupervised,flaschel2023automated} formulates the discovery task as a sparse regression problem on model libraries constructed from, e.g., Ogden or generalized Mooney-Rivlin features to describe hyperelastic materials, directly enforcing equilibrium in weak form. This regression-based perspective enables the automated selection of parsimonious closed-form constitutive models. Bayesian extensions of EUCLID \cite{joshi2022bayesian} further quantify uncertainty, while neural-network-based formulations \cite{thakolkaran2022nn,lourencco2024indirect,li2022equilibriumCNN} have demonstrated the potential of unsupervised deep learning for constitutive identification. Variational System Identification (VSI) \cite{wang2021inference,wang2021variational}, also inspired by VFM, offers a different weak-form strategy, where global sparse regression is replaced with stepwise elimination based on statistical hypothesis testing, thereby pruning insignificant terms from the candidate set. \Rev{Recently, a database-driven strategy called Material Fingerprinting (MF)~\cite{flaschel2025material} was presented. MF reinterprets the discovery problem as a pattern recognition task, matching experimental data against a precomputed database of stress responses from a finite set of established constitutive models (e.g., Gent, Ogden). This avoids online optimization, but comes at the cost of constructing a comprehensive database.}

\Rev{
	While a recent study \cite{abbasi2025discovery} has demonstrated that EUCLID can be applied to experimental full-field displacement data, VFM-based approaches inherently require dense spatial data to compute deformation gradients. They cannot be applied directly to sparse point measurements (e.g., only a few sensors).}
The practical deployment of unsupervised approaches is further constrained by data quality. Noise amplification during numerical or automatic differentiation is well known to degrade parameter identification \cite{avril2004sensitivity,rossi2015effect}.
To improve robustness, some recent discovery frameworks have introduced Bayesian sparsity priors \cite{hirsh2022sparsifying} or ensemble-based updating strategies \cite{fasel2022ensemble}, which can stabilize the identification of partial differential equations (PDEs) from noisy or incomplete data. While promising in the context of general PDEs identification, these strategies do not address the mechanics-specific challenge of constitutive model discovery.

A more systematic treatment of uncertainty has long been established in the uncertainty quantification community, where iterative Bayesian assimilation frameworks are used to deal with imperfect data. By treating model calibration as a sequential inference problem, such approaches refine predictions by continuously updating probabilistic beliefs in light of noisy observations \cite{kennedy2001bayesian,jiang2015surrogate,Arendt2012}. The {statistical finite element method} (statFEM) \cite{girolami2021statistical,akyildiz2022statistical,koh2023stochastic,narouie2023inferring} is a prominent example: it reconstructs full-field displacements from sparse and noisy measurements by applying Bayesian inference directly into the finite element solution. Recent works have established polynomial convergence guarantees \cite{karvonen2025error} and probabilistic displacement reconstructions under polynomial chaos expansions (PCE) \cite{narouie2025mechanical}. Yet, while statFEM robustly recovers displacement fields under uncertainty, it neither aims to strictly enforce constitutive consistency nor to provide interpretable material models.



In this work, we propose an iterative framework that couples statFEM with VFM and EUCLID, combining the strengths of both approaches. The resulting framework, statFEM--EUCLID, integrates the probabilistic assimilation of sparse, noisy measurements with automated discovery of constitutive laws. Rather than applying EUCLID directly to noisy, sparse displacement fields, we first assimilate the measurements through statFEM, thereby filtering noise and reconstructing missing data in a probabilistically consistent manner. The assimilated fields are then used as input to EUCLID, which performs automated discovery of constitutive models within a model library. We demonstrate that this approach extends the applicability of unsupervised model discovery to regimes of severe sparsity in the presence of noise.

The remainder of this manuscript is organized as follows. In \autoref{sec:PaperC_MDisc}, the proposed statFEM--EUCLID framework is developed from rigorous theory, following the notion of reduced, intermediate, and all-at-once approaches for parameter identification and discovery \cite{romer2025reduced}. \autoref{sec:PaperC_Forward_Modeling} presents the governing equations of the boundary value problem, the construction of the constitutive model library, and the PCE used for uncertainty propagation, i.e., the computation of forecasted displacements. The forecasted, stochastic displacement field is assimilated with sparse, noisy measurements as detailed in \autoref{sec:PaperC_Statistical_Inference}, and the subsequent discovery step is outlined in \autoref{sec:euclid}. Numerical examples in \autoref{sec:PaperC_NumExamples} demonstrate the effectiveness of the framework for hyperelastic model discovery from sparse and noisy data. Finally, conclusions and perspectives are provided in \autoref{sec:PaperC_Conclusion_Outlook}.

\section{Unsupervised Constitutive Model Discovery} 
\label{sec:PaperC_MDisc}

\Rev{Let $\vecu \in \mathbb{R}^{\ngdof}$ denote the displacement vector and $\veckappa\in\mathbb{R}^{n_{\phi}}$ a vector of material parameters. The inverse problem is governed by the state and observation equations:}
\begin{align}
	 & \pmb{\mathcal{F}}(\vecu \, , \, \veckappa) = \veczero , & \label{eq:PaperC_state_eq} \\
	 & \vecy =  \Rev{\matH} \, \vecu + \vece.                  & \label{eq:PaperC_obs_eq}
\end{align}
\Rev{Here, \(\pmb{\mathcal{F}} \in \mathbb{R}^{\ngdof} \) is the residual of the governing equations of a boundary value problem (c.f.~\autoref{sec:PaperC_Boundary_Value_Problem} and \autoref{sec:euclid}). The observation operator $\matH\in\mathbb{R}^{\ngsen\times\ngdof}$ is a mapping from FE nodes to sensor locations, c.f. \cite{girolami2021statistical, narouie2023inferring, narouie2025mechanical} for details, $\vecy\in\mathbb{R}^{\ngsen}$ is the observation vector and \( \vece \in \mathbb{R}^{\ngsen}\) a noise vector, typically modeled according to the specifications of the sensing system. With $\nsen$ the number of sensors, $\dof$ the number of degrees of freedom and $d$ the physical dimension of the problem domain, we define  $\ngsen = \nsen \times d$  and $\ngdof = \dof \times d$.}

In a deterministic setting, we can identify both the unknown displacements $\vecu$ and material parameters $\veckappa$ such that the $\ell_2$ norms of state \eqref{eq:PaperC_state_eq} and observation \eqref{eq:PaperC_obs_eq} are minimized, c.f.~\cite{romer2025reduced}:
\begin{equation}\label{eq:PaperC_coupled_eq}
	\{\vecu^*, \veckappa^*\} = \arg\min_{\vecu, \veckappa} \left[ \omega_s \, \left\| \pmb{\mathcal{F}}(\vecu \, , \, \veckappa) \right\|_2^2 + \omega_e  \, \left\| \Rev{\matH} \, \vecu - \vecy \right\|_2^2 \right].
\end{equation}
Here, the noise is considered negligible ($\vece=\vec0)$ and \( \omega_s \) and \( \omega_e \) control the relative importance between physics (state equation) and data (observation equation).
Note that, along the same lines, the state and observation equations \eqref{eq:PaperC_state_eq} and \eqref{eq:PaperC_obs_eq} can be considered as stochastic. In this case, under suitable assumptions, \eqref{eq:PaperC_coupled_eq} can be derived from a Maximum a Posteriori (MAP) estimate, see \autoref{sec:map} for details.

The problem stated in \eqref{eq:PaperC_coupled_eq} is referred to as the all-at-once approach (AAO) and is rarely used in practice. \Rev{Instead, the reduced approach with $\omega_s\rightarrow\infty$ is most frequently employed, i.e. the state equation is treated as exact. The FEMU approach falls under this category, see e.g. \cite{romer2025reduced} and references therein. Herein, we employ an intermediate approach with $\omega_e \to \infty$, where the observation equation is treated as exact. This concept is employed in the VFM \cite{grediac2006virtual, pierron2012virtual}.}
\Rev{Here,} the state equation  is minimized directly for the material parameters, and we obtain:
\begin{equation}\label{eq:VFM}
	\veckappa^* = \arg\min_{\veckappa} \left\| \pmb{\mathcal{F}}(\vecy \, , \, \veckappa) \right\|_2^2 \quad \text{subject to} \, \, \Rev{\matH} \, \vecu - \vecy=\vec0.
\end{equation}
In the context of constitutive model discovery, one seeks to identify a sparse parameter vector $\veckappa^*\in\mathbb{R}^{n_{\phi}^{\mathcal{A}}}$, with $n_{\phi}^{\mathcal{A}} << n_{\phi}$. Sparsity can be enforced, for instance, using $\ell_p$-regularization, which has been promoted in the context of EUCLID \cite{flaschel2021unsupervised}. In the present contribution, we consider an $\ell_1$- or LASSO-regularized \cite{mcculloch2024LPRegularizationModelDiscovery} version of the VFM \eqref{eq:VFM}, which yields:
\begin{equation}
	\veckappa_{\lambda} = \arg\min_{\veckappa} \left\| \pmb{\mathcal{F}}(\vecy \, , \, \veckappa) \right\|_2^2 + \lambda \|\veckappa\|_1 \quad \text{subject to} \, \, \Rev{\matH} \, \vecu - \vecy=\vec0,
	\label{eq:euclid_1}
\end{equation}
where $\lambda>0$ is a hyperparameter. $\veckappa^*$ is selected such that $\veckappa^* = \veckappa_{\lambda^*}$, where $\lambda^*$ yields the most parsimonious model that can explain the observed data. Further details on constitutive model discovery with EUCLID are provided in~\autoref{sec:euclid}.

\begin{figure}[!htb]
	\centering
	\includegraphics{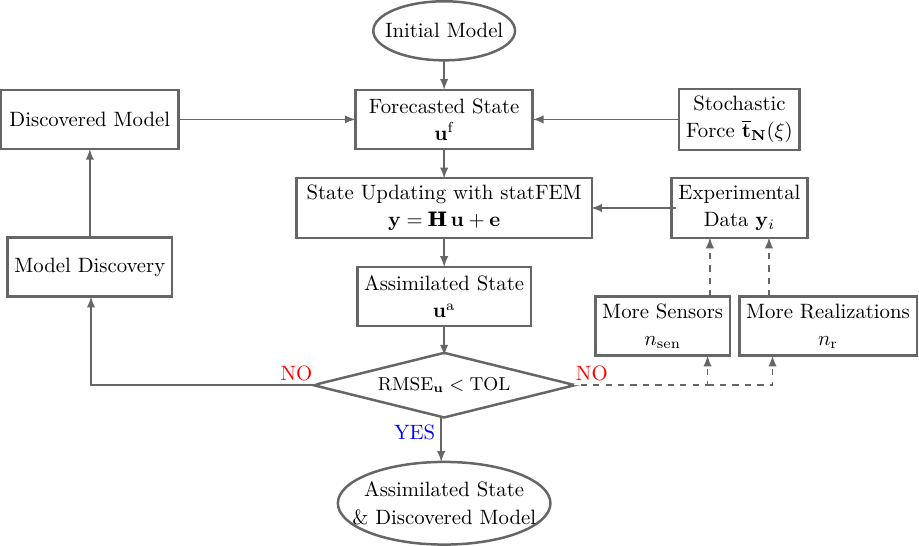}
	\caption{Graphical illustration of the iterative model discovery framework. The loop iteratively updates the initial model predictions based on observational data via statFEM and discovers a constitutive model until convergence is reached. The dashed line shows the displacement assimilation process without updating the constitutive model. This framework is adapted from \cite{Arendt2012}.}
	\label{fig:graphical_illustration}
\end{figure}

Both VFM \eqref{eq:VFM} and EUCLID \eqref{eq:euclid_1} rely on the availability of full-field data $\vecy$. To address this limitation, we propose an iterative framework based on stochastic forward modeling, Bayesian inference, and sparse regression. The complete pipeline is summarized in \autoref{fig:graphical_illustration} and proceeds as follows:

\begin{enumerate}
	\item \textbf{Uncertainty propagation:}
	      For given material parameters $\veckappa$ and uncertain boundary traction \( \overline{\tent}_{\text{N}}(\xi) \), a stochastic boundary value problem is solved (see~\autoref{sec:PaperC_Forward_Modeling}). Its solution is referred to as the \Rev{nodal} forecasted displacement field \Rev{\( \vecu^{\text{f}} \in \mathbb{R}^{\ngdof} \)} that follows from the state equation:
	      \begin{equation}
		      \pmb{\mathcal{F}}( \vecu^{\text{f}} \, , \, \veckappa) = \veczero, \quad \text{where} \, \, \vecu^{\text{f}} \sim f_{\vecu^{\text{f}}}(\vecu^{\text{f}})  = \mathcal{N}(\vecmu_{\vecu^{\text{f}}}, \matC_{\vecu^{\text{f}}}).
		      \label{eq:PaperC_forward_eq}
	      \end{equation}

	\item \textbf{Bayesian state update (statFEM):}
	      Given the observational data \( \vecy \) and the forecast \( \vecu^{\text{f}}\), the observation model~\eqref{eq:PaperC_obs_eq} is used to perform a Bayesian update of the displacement field using \mbox{statFEM.}
	      \Rev{The assimilated nodal state \( \vecu^{\text{a}} \in \mathbb{R}^{\ngdof} \), with posterior density \( f_{\vecu^{\text{a}}} \),
		      is determined from the observation equation~\eqref{eq:PaperC_obs_eq}, yielding:}
	      \begin{equation}
		      \vecu^{\text{a}}\sim f_{\vecu^{\text{a}}}(\vecu^{\text{a}})  = \mathcal{N}(\vecmu_{\vecu^{\text{a}}}, \matC_{\vecu^{\text{a}}}).
		      \label{eq:PaperC_posterior_pdf}
	      \end{equation}
	      Here, \( \vecmu_{\vecu^{\text{a}}}  \Rev{\in \mathbb{R}^{\ngdof}} \) and \( \matC_{\vecu^{\text{a}}} \Rev{\in \mathbb{R}^{\ngdof \times \ngdof}}\) denote the posterior mean and covariance of the assimilated state, see~\autoref{sec:PaperC_Statistical_Inference} for details.

	\item \textbf{Convergence check:}
	      The Root Mean Squared Error (RMSE) between assimilated state \( \vecmu_{\vecz} = \Rev{\matH} \, \vecmu_{\vecu^{\text{a}}} \) at sensor locations and data $\vecy$ is defined as:
	      \begin{equation}
		      \text{RMSE}_{\vecu} =  \sqrt{ \frac{ \norm{ \vecmu_{\vecz} - \vecy}_2^2 }{ \nsen } }.
		      \label{eq:PaperC_rmse}
	      \end{equation}
	      Convergence is defined by thresholding:
	      \begin{equation}
		      \text{RMSE}_{\vecu} < \text{TOL}
	      \end{equation}
	      We continue with Step~4 until convergence is reached.

	\item \textbf{statFEM-VFM / statFEM-EUCLID:}
	      Material parameters \( \veckappa \) are estimated via VFM \eqref{eq:VFM} or EUCLID \eqref{eq:euclid_1}, subject to the stochastic observation equation:
	      \begin{equation}
		      \begin{aligned}
			      \veckappa_{\lambda} & = \arg\min_{\veckappa} \left\| \pmb{\mathcal{F}}(\vecmu_{\vecu^{\text{a}}} \, , \, \veckappa) \right\|_2^2 + \lambda \|\veckappa\|_1 \quad \text{subject to} \, \,
			      \Rev{\matH} \, \vecmu_{\vecu^{\text{a}}} + \vece- \vecy = \vec0.
			      \label{eq:PaperC_generic_discovery}
		      \end{aligned}
	      \end{equation}
	      {The assimilated mean displacement $\vecmu_{\vecu^{\text{a}}}$ is considered the approximate solution of the stochastic observation equation \eqref{eq:PaperC_obs_eq}. This viewpoint allows us to equate the state $\vecu$ with the assimilated mean ($\vecu = \vecmu_{\vecu^{\text{a}}}$) in the deterministic VFM or EUCLID formulations.} After this step, we go back to Step~1 using $\veckappa^*=\veckappa_{\lambda^*}$, see   \autoref{sec:euclid}.

\end{enumerate}
If convergence cannot be reached, additional data can be introduced to improve identifiability and robustness - either by increasing the number of sensors \( n_{\text{sen}} \), or by increasing the number of realizations per sensor \( n_{\text{r}} \). The effect of  $n_{\text{sen}}$ and $n_{\text{r}}$ has been studied, e.g., in \cite{girolami2021statistical, narouie2023inferring, narouie2025mechanical}. In this contribution, we restrict ourselves to cases with $n_{\text{r}} = 1$ and examine the effect of $n_{\text{sen}}.$

\section{Forward Problem and Uncertainty Propagation}
\label{sec:PaperC_Forward_Modeling}

Throughout this work, we consider the balance of linear momentum subject to Dirichlet and Neumann boundary conditions as the state equation. Both the deterministic and stochastic boundary value problems are defined in \autoref{sec:PaperC_Boundary_Value_Problem}. These problems are complemented by a constitutive model library introduced in \autoref{sec:PaperC_constitutive_library}. In \autoref{sec:PaperC_pce}, the \Rev{Polynomial-Chaos}-based surrogate model of the stochastic boundary value problem is discussed.

\subsection{Boundary Value Problem}
\label{sec:PaperC_Boundary_Value_Problem}

The deterministic boundary value problem (DBVP) seeks the displacement $\vecu$ such that:
\begin{equation}
	\begin{cases}
		\nabla \cdot \tenP(\nabla \vecu; \veckappa) + \vecb = \veczero,              & \text{in } \physicalDomain, \\[1ex]
		\vecu - \preStochDisp = \vec0,                                               & \text{on } \preDispOnBound, \\[1ex]
		\tenP(\nabla \vecu; \veckappa) \cdot \unitNormal - \overline{\tent} = \vec0, & \text{on } \preTracOnBound.
	\end{cases}
	\label{eq:PaperC_DBVP_general}
\end{equation}
\Rev{Here, with a slight abuse of notation, \( \vecu\in\mathbb{R}^3 \)  denotes the displacement field rather than the nodal displacement vector.} The abstract formulation \eqref{eq:PaperC_DBVP_general} accommodates various types of constitutive behavior. Herein, we consider a linear elastic model
\begin{equation}
	\tenP(\nabla \vecu; \veckappa) = \mathbb{C} : \boldsymbol{\varepsilon},
\end{equation}
with \( \boldsymbol{\varepsilon} = \tfrac{1}{2}(\nabla \vecu + \nabla \vecu^\top) \) the linear strain measure and $\mathbb{C}$ the isotropic elasticity tensor defined by Young's modulus $E$ and Poisson ratio $\nu$. In addition, we consider hyperelastic models from a library introduced in \autoref{sec:PaperC_constitutive_library}. The discretization of the DBVP \Rev{yields the residual $\pmb{\mathcal{F}}$ as} discussed in the context of EUCLID in \autoref{sec:euclid}.

In forward simulations involving uncertainty, we consider stochasticity only in the external loading, not in the material model or its parameters. \Rev{Thus, the traction \( \overline{\tent}(\xi) \) is parametrized with a standard Gaussian random variable \( \xi \).} This yields the stochastic boundary value problem (SBVP):
\begin{equation}
	\begin{cases}
		\nabla \cdot \tenP(\nabla \stochDisp(\xi); \veckappa) + \vecb = \veczero,                     & \text{in } \physicalDomain, \\[1ex]
		\stochDisp( \xi) - \preStochDisp = \vec0,                                                     & \text{on } \preDispOnBound, \\[1ex]
		\tenP(\nabla \stochDisp( \xi); \veckappa) \cdot \unitNormal - \overline{\tent}( \xi) = \vec0, & \text{on } \preTracOnBound.
	\end{cases}
	\label{eq:PaperC_SBVP_general}
\end{equation}
Notice that the DBVP \eqref{eq:PaperC_DBVP_general} leads to the VFM, whereas the SBVP \eqref{eq:PaperC_SBVP_general} is required in the statFEM step to compute the forecasted displacement. Herein, the solution of the SBVP is approximated via PCE, see \autoref{sec:PaperC_pce}.

\subsection{Constitutive Model Library}
\label{sec:PaperC_constitutive_library}

The internal mechanical response of hyperelastic materials is characterized by the first Piola--Kirchhoff stress tensor \( \tenP \), which is derived from a strain energy density function \( W \). To enable model discovery, we express the strain energy density function in a series expansion of pre-defined basis functions $\{\phi_j\}_{j=1}^{n_\phi}$:
\begin{equation}
	W(\tenF;\veckappa) = \sum_{j=1}^{n_\phi} \kappa_j\, \phi_j(\tenF).
	\label{eq:PaperC_w_series_expansion}
\end{equation}
Following standard thermodynamic arguments, the first Piola--Kirchhoff stress tensor $\tenP$ is obtained from:
\begin{equation}\label{eq:pk1_1}
	\tenP = \frac{\partial W(\tenF;\veckappa)}{\partial \tenF} = \sum_{j=1}^{n_\phi} \kappa_j \, \frac{\partial \phi_j}{\partial \tenF}.
\end{equation}
We split the strain energy into an isochoric and a volumetric component:
\begin{equation}
	W(J_1, J_2, J_3) = W_{\text{iso}}(J_1, J_2) + W_{\text{vol}}(J_3).
	\label{eq:PaperC_W_series_expansion}
\end{equation}
The energy density is defined in terms of modified invariants of the right Cauchy--Green tensor \( \matC = \defGrad^\top \defGrad \). Specifically, we use:
\begin{equation}
	J_1 = I_1 I_3^{-1/3}, \qquad J_2 = I_2 I_3^{-2/3}, \qquad J_3 = I_3^{1/2},
\end{equation}
where \( I_1, I_2, I_3 \) are the standard invariants of \( \matC \), see~\cite{kim2014introduction}. The isochoric component is modeled as a polynomial expansion:
\begin{equation}
	W_{\text{iso}}(J_1, J_2) = \sum_{\substack{i + j = 1  \\ i,j \geq 0}}^{N_{\text{MR}}} A_{ij} \, \phi_{ij}^{\text{iso}}(J_1, J_2) = \sum_{\substack{i + j = 1  \\ i,j \geq 0}}^{N_{\text{MR}}} A_{ij} \left( J_1 - 3 \right)^i \left( J_2 - 3 \right)^j,
	\label{eq:PaperC_W_iso_extended}
\end{equation}
where $\phi_{ij}^{\text{iso}}(J_1, J_2)$ are isochoric features and \( A_{ij} \) are the isochoric material parameters.
The volumetric part is expanded in even powers of \( (J_3 - 1) \):
\begin{equation}
	W_{\text{vol}}(J_3) = \sum_{i=1}^{N_{\text{vol}}} \, B_i \, \phi_{i}^{\text{vol}}(J_3) =  \sum_{i=1}^{N_{\text{vol}}}  \, B_i \left( J_3 - 1 \right)^{2i}.
	\label{eq:PaperC_W_vol_extended}
\end{equation}
Here, $\phi_{i}^{\text{vol}}(J_3)$ are volumetric features and \( B_i \) are the volumetric material parameters. The full vector of material parameters is defined as
\begin{equation}
	\veckappa :=
	\left[ A_{10}, A_{01}, \hdots, {A_{0N_{\text{MR}}}}, B_{1}, \hdots,  B_{N_{\text{vol}}}\right] \in\mathbb{R}^{n_{\phi}}.
	\label{eq:PaperC_veckappa}
\end{equation}
The complete list of included basis functions and their corresponding labels is provided in~\autoref{sec:PaperC_basis_functions}.

\subsection{Polynomial Chaos Expansion for Forecasted Displacements}
\label{sec:PaperC_pce}

In this work, uncertainty in forecasted displacements is introduced exclusively through the applied boundary tractions. \Rev{This choice provides a simple stochastic prior for the statFEM update and keeps the formulation as lightweight as possible. The proposed framework does not rely on this specific assumption. Uncertainty may be introduced in the material parameters, in the external loading, or in both. In the present study we focus on the simplest stochastic setting in order to demonstrate the iterative reconstruction and discovery procedure. More general stochastic formulations can be incorporated in future work.} The traction field \( \overline{\tent}( \xi) \) from \eqref{eq:PaperC_SBVP_general} is modeled as a random field with stochastic dependence on a scalar Gaussian random variable \( \xi \sim \mathcal{N}(0, 1) \). Specifically, we define:
\begin{equation}
	\overline{\tent}(\xi) = \vecmu_{\overline{\tent}} + \vecsigma_{\overline{\tent}} \, \xi,
	\label{eq:PaperC_traction_random}
\end{equation}
where \( \vecmu_{\overline{\tent}}\) denotes the deterministic mean traction vector and \( \vecsigma_{\overline{\tent}} \) is the standard deviation vector. To propagate this uncertainty through the forward problem, there exist intrusive \cite{deb2001solution,matthies2005galerkin} and non-intrusive approaches \cite{berveiller2006stochastic,reagana2003uncertainty}. We employ a non-intrusive PCE to approximate the stochastic displacement field. The forecasted (prior) displacement \( \vecu^{\text{f}}(\xi) \) is then expressed as:
\begin{equation}
	\vecu^{\text{f}}(\xi) \approx \sum_{j=0}^{P_u} \vecu_{j} \, \Psi_{j}(\xi),
	\label{eq:PaperC_U_PC}
\end{equation}
where \( \vecu_{j} \in \mathbb{R}^{\ngdof} \) are deterministic PC coefficients with \( \ngdof = \dof \cdot d \) the total number of degrees of freedom and $d$ the number of dimensions. Moreover, \( \Psi_{j}(\xi) \) are Hermite polynomials, and \( P_u \) is the total number of modes in the expansion.
\Rev{The use of a PCE surrogate reduces the computational cost of the Bayesian update to negligible levels during the iterative discovery process, as no  finite element solves are required in the data-assimilation step.}

In this contribution, the stochastic regression method is chosen to identify the PCE coefficients \( \vecu_{j} \)~\cite{berveiller2005non}. Given \( S \) Monte Carlo samples \( \{ \xi_s \}_{s=1}^S \) and the corresponding finite element solutions \( \vecu( \xi_s) \), the coefficients are computed by minimizing the mean squared error between the actual displacements and their PCE surrogate:
\begin{equation}
	(\vecu_{0}, \ldots, \vecu_{P_u}) =
	\argmin_{\vecu_{j}} \frac{1}{S} \sum_{s=1}^{S}
	\left\| \vecu( \xi_s) - \sum_{j=0}^{P_u} \vecu_{j}\, \Psi_{j}(\xi_s) \right\|_2^2.
	\label{eq:PaperC_PC_minimization}
\end{equation}
This surrogate model enables efficient evaluation of statistical quantities. In particular, the mean and covariance of the forecasted displacement field are given by
\begin{equation}
	\vecmu_{\vecu^{\text{f}}} = \vecu_{0}, \qquad
	\matC_{\vecu^{\text{f}}} = \sum_{j=1}^{P_u} \langle (\Psi_{j})^2 \rangle \, \vecu_{j} \, (\vecu_{j})^\top,
	\label{eq:PaperC_meanCovPC}
\end{equation}
where \( \langle (\Psi_{j})^2 \rangle \) denotes the expected value of the squared basis function. Note that because the single random variable $\xi$, the covariance matrix $\matC_{\vecu^{\text{f}}}$ exhibits nonzero off-diagonal terms, which reflects spatial correlation of the displacement field. The PCE approximation \( \vecu^{\text{f}}(\xi) \) serves as the prior displacement field in the subsequent Bayesian update, where it is assimilated with noisy sensor observations through the statFEM, see~\autoref{sec:PaperC_Statistical_Inference}.

\section{Statistical Inference with statFEM}
\label{sec:PaperC_Statistical_Inference}

In statFEM, the observation equation~\eqref{eq:PaperC_obs_eq} is treated as a statistical generating model, and the goal is to compute the posterior distribution of the displacement field given the observed sensor data. By Bayes' rule, the posterior density of the displacement field conditioned on the observed measurements is given by:
\begin{equation}
	f_{\mathbf{u} | \vecy}(\stochDisp \mid \vecy) = f_{\mathbf{u}^{\text{a}}}(\stochDisp^{\text{a}}) = \frac{\PDF{\vecy | \mathbf{u}^{\text{f}}}{\vecy \, | \, \stochDisp^{\text{f}}} \, \PDF{\mathbf{u}^{\text{f}}}{\stochDisp^{\text{f}}}}{\PDF{\vecy}{\vecy}}.
\end{equation}
The likelihood $f_{\vecy | \mathbf{u}^{\text{f}}}$ is obtained by evaluating the discretized observation equation \eqref{eq:PaperC_obs_eq}, defined by
\begin{equation}
	\vecy = \matH \, \vecu^{\mathrm{f}} + \vece.
\end{equation}
We assume that the forecasted displacement $\mathbf{u}^{\text{f}}$ follows a Gaussian distribution obtained from the PCE (\autoref{sec:PaperC_pce}):
\begin{equation}
	\mathbf{u}^{\text{f}} \sim \mathcal{N}(\vecmu_{\vecu^{\text{f}}}, \matC_{\vecu^{\text{f}}}).
\end{equation}
Measurement noise $\vece$ is modeled as a Gaussian Process evaluated at $n_{\text{sen}}$ sensor locations, resulting in:
\begin{equation}
	\vece \sim \mathcal{N}(\mathbf{0}, \matC_{\vece}), \quad \matC_{\vece} = \sigma_{\vece}^2 \, \, \matI.
	\label{eq:PaperC_ei_repeated}
\end{equation}
Assuming $n_{\text{r}}$ independent sensor readings, the posterior distribution of the displacement field is again Gaussian, with closed-form expressions for the mean and covariance:
\begin{align}
	\vecmu_{\vecu^{\text{a}}} & = \matC_{\vecu^{\text{a}}} \left( \matH^{\top} \, \matC_{\vece}^{-1}  \, \sum_{i=1}^{n_{\text{r}}} \vecy_{i}  \, +  \, \Big( \operatorname{SPD}(\matC_{\vecu^{\text{f}}}) \Big)^{-1} \vecmu_{\vecu^{\text{f}}} \right),
	\label{eq:PaperC_post_mean_repeated}                                                                                                                                                                                                                \\
	\matC_{\vecu^{\text{a}}}  & = \left( n_{\text{r}}  \, \matH^{\top}  \, \matC_{\vece}^{-1}  \, \matH  \, +  \, \Big( \operatorname{SPD}(\matC_{\vecu^{\text{f}}}) \Big)^{-1} \right)^{-1}.
	\label{eq:PaperC_post_cov_repeated}
\end{align}
Usually the covariance matrix $\matC_{\vecu^{\text{f}}}$ given in \eqref{eq:PaperC_meanCovPC} is ill-conditioned, therefore we use the $\operatorname{SPD}(\matC_{\vecu^{\text{f}}})$ instead of $\matC_{\vecu^{\text{f}}}$. The operator $\operatorname{SPD}(\bullet)$ denotes the nearest symmetric positive definite matrix; see~\autoref{sec:PaperC_nearest_spd}.
Finally, the distribution of the predicted displacement at sensor locations is given by:
\begin{equation}
	\vecz \sim \mathcal{N}(\vecmu_{\vecz}, \matC_{\vecz}), \quad
	\vecmu_{\vecz} = \matH \, \vecmu_{\vecu^{\text{a}}}, \quad
	\matC_{\vecz} = \matH \, \matC_{\vecu^{\text{a}}} \, \matH^\top.
\end{equation}

\section{statFEM-EUCLID}\label{sec:euclid}

Let $\vecv$ be a kinematically admissible virtual displacement with $\vecv = \vec0$ on $\partial\Omega_{\vecu}$. Multiplying the equilibrium equation from \eqref{eq:PaperC_DBVP_general} with $\vecv$ and integrating over $\Omega$, then applying the divergence theorem and the boundary conditions, we obtain:
\begin{align}
	\Rev{\int_\Omega \tenP(\nabla \vecu; \veckappa) : \nabla \vecv \, \mathrm{d}\Omega
		- \int_{\partial\Omega_{\tenP}} \overline{\vect}_{\!N} \cdot \vecv \, \mathrm{d}S
		=\vec0.}
	\label{eq:weak_form_general}
\end{align}
Substituting the constitutive model \eqref{eq:pk1_1} into the weak form \eqref{eq:weak_form_general}, and discretizing the domain using finite elements, we obtain:
\begin{equation}\label{eq:feature_matrix}
	\Rev{\pmb{\mathcal{F}}(\vecu, \veckappa) =\matA(\vecu) \, \veckappa - \vecp=\vec0.}
\end{equation}
Here, $\matA(\vecu) \in \mathbb{R}^{n_{\mathrm{gdof}} \times n_\phi}$ is the feature matrix and $\vecp \in \mathbb{R}^{n_{\mathrm{gdof}}}$ is the assembled external force vector. With \eqref{eq:feature_matrix}, the EUCLID formulation  from \eqref{eq:euclid_1} becomes:
\begin{equation}
	\veckappa_{\lambda} = \arg\min_{\veckappa \geq \vec0} \left\| \matA(\vecy) \veckappa - \vecp \right\|_2^2 + \lambda \mathbf{1}^\top \veckappa \quad \text{subject to} \, \, \matH \, \vecu- \vecy = \vec0.
	\label{eq:PaperC_euclid_final}
\end{equation}
Here, the regularization term is written as \( \lambda \mathbf{1}^\top \veckappa \), which is equivalent to the standard LASSO penalty \( \lambda \|\veckappa\|_{p=1} \) for $\veckappa\geq \vec0$.
\Rev{The non-negativity constraint $\veckappa \geq \vec0$ is imposed as a sufficient condition for physical admissibility. Since the basis functions in the library are chosen to be polyconvex with respect to the deformation measures, enforcing non-negative coefficients ensures that the identified total strain energy density preserves these stability properties \cite{hartmann_generalizedStrainEnergyFunctions_2003}.}
We note, however, that this is not a strictly necessary condition---other sign patterns may still be physically consistent depending on the form of the basis functions and the specific material behavior under consideration \cite{ricker_systematicFittingHyperelasticity_2023}.

\begin{figure}[!htb]
	\centering
	\subfloat[Pareto analysis full path]{
		\centering
		\includegraphics{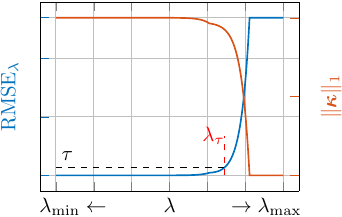}
		\label{fig:pareto_full}
	}
	\hspace*{1cm}
	\subfloat[Pareto analysis for $\text{RMSE}_{\lambda} < \tau$]{
		\centering
		\includegraphics{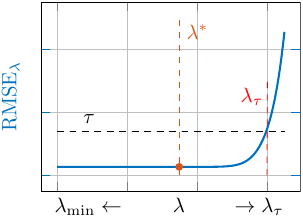}
		\label{fig:pareto_fullpath_RMSE70}
	}
	\caption{Pareto analysis of the model discovery problem~\eqref{eq:PaperC_discovery_exntedned}. (a) Full path. (b) Zoomed view for $\text{RMSE}_{\lambda} < \tau$.}
	\label{fig:pareto_front}
\end{figure}

In our statFEM-EUCLID formulation, the displacement field is equated with the mean of the assimilated state $\vecmu_{\vecu^{\text{a}}}$. Under this assumption and making use of \eqref{eq:feature_matrix},  the model discovery problem first stated in \eqref{eq:PaperC_generic_discovery} reduces to:
\begin{equation}
	\veckappa_{\lambda} = \arg\min_{\veckappa \geq \vec0} \left\| \matA(\vecmu_{\vecu^{\text{a}}}) \cdot \veckappa - \vecp \right\|_2^2 + \lambda \mathbf{1}^\top \veckappa \quad \text{subject to} \, \,
	\tenH \, \vecmu_{\vecu^{\text{a}}} + \vece- \vecy = \vec0.
	\label{eq:PaperC_discovery_linear}
\end{equation}
Additionally, a volumetric penalty constraint of the form
\begin{equation}
	\sum_{\substack{i + j = 1  \\ i,j \geq 0}}^{N_{\text{MR}}} A_{ij} - r \, \sum_{i=1}^{N_{\text{vol}}} \, B_i = 0,
\end{equation}
is enforced, where $r > 0$ is a prescribed penalty multiplier (in our implementation, $r = 3$).  This constraint prevents unrealistically large volumetric terms relative to the isochoric contributions. Therefore, the model discovery problem stated in \eqref{eq:PaperC_discovery_linear} is extended with the volumetric penalty constraint as follows:
\begin{equation}
	\veckappa_{\lambda} = \arg\min_{\veckappa \geq \vec0} \left\| \matA(\vecmu_{\vecu^{\text{a}}}) \cdot \veckappa - \vecp \right\|_2^2 + \lambda \mathbf{1}^\top \veckappa \quad \text{subject to} \, \,
	\begin{cases}
		\tenH \, \vecmu_{\vecu^{\text{a}}} + \vece- \vecy = \vec0 \\
		\sum_{\substack{i + j = 1                                 \\ i,j \geq 0}}^{N_{\text{MR}}} A_{ij} - r \, \sum_{i=1}^{N_{\text{vol}}} \, B_i = 0.
	\end{cases}
	\label{eq:PaperC_discovery_exntedned}
\end{equation}
The solution of~\eqref{eq:PaperC_discovery_exntedned} depends sensitively on the choice of the regularization parameter~$\lambda$. To quantify the trade-off between accuracy and sparsity, we perform a Pareto analysis over the range
\begin{equation}\label{eq:lambda_min_max}
	\lambda \in \mathrm{logspace}(\lambda_{\min}, \lambda_{\max}, N_\lambda),
\end{equation}
and for each $\lambda$ solve the optimization problem~\eqref{eq:PaperC_discovery_exntedned}. The residual error $\text{RMSE}_{\lambda}$ of the resulting model is measured by
\begin{equation}
	\text{RMSE}_{\lambda}
	= \sqrt{\frac{\left\| \vecp - \matA \, \veckappa_{\lambda} \right\|_2^2}{n_{\mathrm{gdof}}} },
	\label{eq:PaperC_RMSE_lambda}
\end{equation}
where $\veckappa_{\lambda}$ denotes the coefficient vector obtained at~$\lambda$.
The outcome of a Pareto analysis is illustrated in~\autoref{fig:pareto_full}.  As can be seen, increasing~$\lambda$ strengthens the regularization, which monotonically decreases the $\ell_{1}$-norm of $\veckappa_{\lambda}$, while simultaneously degrading the fit accuracy, i.e. $\mathrm{RMSE}_{\lambda}$.
In addition, we define the number of active terms $n^{\mathcal{A}}_{\lambda} \subseteq n_\phi$ of the model at~$\lambda$ as
\begin{equation}
	n^{\mathcal{A}}_{\lambda} := \#\left\{ j \;\middle|\; \kappa_{j} \in \veckappa_{\lambda}, \,  \kappa_{j} > 10^{-10} \right\},
	\label{eq:PaperC_active_coeff}
\end{equation}
that is, $n^{\mathcal{A}}_{\lambda}$ counts how many coefficients of the solution vector $\veckappa_\lambda$ are greater than $10^{-10}$ at a specific $\lambda$ such that the rest can be assumed to equal zero numerically. Additionally, we define the admissible set of the penalty terms as
\begin{equation}\label{eq:lambda_acc}
	\Lambda_{\mathrm{acc}} := \left\{ \lambda \in [\lambda_{\min},\,\lambda_{\tau}] \;\middle|\; \mathrm{RMSE}_{\lambda} < \tau \right\},
\end{equation}
where, $\Lambda_{\mathrm{acc}}$ denotes the set of $\lambda$ that pass the admissibility criterion
$\mathrm{RMSE}_{\lambda} < \tau$, with $\tau$ denoting a user-defined tolerance that serves as an admissibility threshold for acceptable residual error, as illustrated in~\autoref{fig:pareto_fullpath_RMSE70}.
In other words, only models whose RMSE falls below the prescribed threshold are retained for further consideration.

Crucially, model selection is not solely about reducing error, but about representing with interpretability and generalization capability~\cite{brunton2022data}. To this end, we first minimize the model complexity, measured by the number of active coefficients $n^{\mathcal{A}}_{\lambda}$, i.e.,
\begin{equation}
	n_{\min}^{\mathcal{A}} = \min_{\lambda \in \Lambda_{\mathrm{acc}}} \; n^{\mathcal{A}}_{\lambda}.
	\label{eq:PaperC_min_active}
\end{equation}
This yields the smallest number of active terms among all admissible models.
Notice that the discovery problem defined in \autoref{sec:PaperC_MDisc} refers to $n^{\mathcal{A}}_\phi$ to indicate the number of active terms relative to the full candidate library of size $n_\phi$. Still, in the EUCLID formulation, the number of active terms is indexed by $\lambda$ and therefore written $n^{\mathcal{A}}_\lambda$ to indicate the path dependency. The corresponding subset of candidates is
\begin{equation}
	\Lambda_{\min} := \left\{ \lambda \in \Lambda_{\mathrm{acc}} \;\middle|\; n^{\mathcal{A}}_{\lambda} = n^{\mathcal{A}}_{\min} \right\}.
	\label{eq:PaperC_lambda_min}
\end{equation}
Among these parsimonious candidates, with $n^{\mathcal{A}}_{\min}$ active terms, we then select the one with the smallest RMSE error,
\begin{equation}
	\lambda^* = \arg\min_{\lambda \in \Lambda_{\min}} \; \text{RMSE}_{\lambda},
	\label{eq:PaperC_final_selection}
\end{equation}
with the corresponding coefficients
\begin{equation}
	\veckappa^* = \veckappa_{\lambda^*}.
\end{equation}

\section{Numerical Examples}
\label{sec:PaperC_NumExamples}

This section demonstrates the performance of the proposed statFEM-EUCLID framework through synthetic numerical experiments.
\Rev{We acknowledge that experimental validation is the ultimate goal; however, in this paper, synthetic data is utilized to allow for an objective quantification of the discovery error against a known ground truth, which is not possible with experimental data where the true constitutive law is unknown.
	Synthetic data allows us to systematically vary noise levels and sensor configurations to verify the algorithm's convergence properties, which is the primary focus of this methodological contribution.}

\begin{figure}[!htb]
	\centering
	\includegraphics{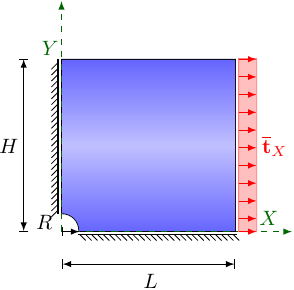}
	\caption{Geometry and loading of the plate with a central hole.}
	\label{fig:Geometry_Loading}
\end{figure}

\begin{figure}[!htb]
	\centering
	\subfloat[Sensor mesh: \( \nsen = 3 \)]{
		\centering
		\includegraphics{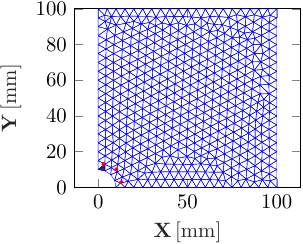}
	}
	\subfloat[Sensor mesh: \( \nsen = 13 \)]{
		\centering
		\includegraphics{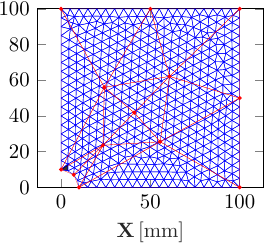}
	}
	\subfloat[Sensor mesh: \( \nsen = 38 \)]{
		\centering
		\includegraphics{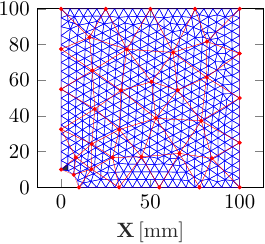}
	}
	\caption{Meshing and sensor locations for different sensor counts. The black element on the bottom left is the representative element, for which we show the strain energy density $W$ in \autoref{fig:plateWithHole_energy_vs_f},  \autoref{fig:plateWithHole_energy_vs_J_e3}, and \autoref{fig:plateWithHole_energy_vs_J_e4}.}
	\label{fig:plateWithHole_mesh}
\end{figure}

\vspace{0.5em}
\noindent \textbf{Synthetic Data Generation.} We consider the plate with a hole depicted in~\autoref{fig:Geometry_Loading} as a case study. Two benchmark materials are used as ground truth: (i) a Neo--Hookean model with one isochoric and one volumetric parameter (\autoref{sec:NH}), and (ii) a Mooney--Rivlin model with two isochoric and one volumetric parameter (\autoref{sec:MR}). The ground-truth displacement fields \( \vecu^{\text{true}} = \vecu(\veckappa^{\text{true}}) \) are computed by solving the deterministic forward FE problems with the true strain energy density $W^{\text{true}}$ and a right-edge traction of the form
\begin{equation}
	\overline{\vect}_{\vecX}
	= \eta\, T_{\max}\, \pmb{\mathcal{E}}_{\vecX},
	\qquad T_{\max}=0.5~\si{MPa},
	\qquad \eta\in[0,1],
\end{equation}
where \(T_{\max}\) is the nominal (maximum) traction magnitude, \(\eta\) a nondimensional load-scaling parameter and \(\pmb{\mathcal{E}}_{\vecX}\) the Cartesian basis vector in the \(X\)-direction. Observations are obtained by projecting the deterministic solution onto the sensor locations and adding Gaussian noise
\begin{equation}
	\vecy = \matH \, \vecu^{\text{true}} + \vece,
	\label{eq:PaperC_displacement_data}
\end{equation}
where \( \vece \sim \mathcal{N}(\veczero, \sigma_{\vece}^2 \, \mathbf{I}) \) is zero-mean Gaussian noise with \( \sigma_{\vece} \in \{10^{-3}, 10^{-4}\} \). We consider three sensor configurations with \( \nsen = \{3, 13, 38\} \), shown in \autoref{fig:plateWithHole_mesh}.

\vspace{0.5em}
\noindent \textbf{Error metrics:} For our synthetic benchmarks, we assess the accuracy of the discovered strain energy density \( W^{\text{disc}} \) w.r.t. the ground truth $W^{\text{true}}$. The relative error $\varepsilon_W$ is defined as
\begin{equation}
	\epsilon_{W} = \frac{1}{n_i n_j n_k} \sum_{i=1}^{n_i} \sum_{j=1}^{n_j} \sum_{k=1}^{n_k}
	\frac{\left[W^{\text{true}}(J_{1i}^{\text{true}}, J_{2j}^{\text{true}}, J_{3k}^{\text{true}}) - W^{\text{disc}}(J_{1i}^{\text{disc}}, J_{2j}^{\text{disc}}, J_{3k}^{\text{disc}})\right]^2}
	{\left[W^{\text{true}}(J_{1i}^{\text{true}}, J_{2j}^{\text{true}}, J_{3k}^{\text{true}})\right]^2},
	\label{eq:PaperC_relative_rmse}
\end{equation}
with $n_i=n_j=n_k=100$ the number of samples.
The sampling ranges are given by the minimum and maximum of observed values:
\begin{equation}
	\begin{aligned}
		 & J_{1i} \in [\min(J_1(\vecu^{\text{true}})), \max(J_1(\vecu^{\text{true}}))], \quad \\
		 & J_{2j} \in [\min(J_2(\vecu^{\text{true}})), \max(J_2(\vecu^{\text{true}}))], \quad \\
		 & J_{3k} \in [\min(J_3(\vecu^{\text{true}})), \max(J_3(\vecu^{\text{true}}))].
	\end{aligned}
\end{equation}
We also define a relative error $\epsilon_{\vecu}$ of the displacement obtained from the discovered model \( \vecu^{\text{disc}} = \vecu(\veckappa_{\lambda^*}) \) w.r.t. the true displacement field \( \vecu^{\text{true}} \):
\begin{equation}\label{eq:PaperC_relative_rmse_u}
	\epsilon_{\vecu} =   \frac{ \norm{ \vecu^{\text{disc}} - \vecu^{\text{true}} }_2 }{\norm{ \vecu^{\text{true}} }_2}.
\end{equation}
Moreover, the pointwise absolute displacement error $\vecvarepsilon_{\vecu}\in\mathbb{R}^{n_{\text{dof}}}$ is defined at each nodal point $\vecX^{(i)}$ as
\begin{equation}\label{eq:point_disp_error}
	\vecvarepsilon_{\vecu} =
	\begin{pmatrix}
		\left| \|\vecu^{\text{disc}}(\vecX^{(1)})\|_2 - \|\vecu^{\text{true}}(\vecX^{(1)})\|_2 \right| \\
		\vdots                                                                                         \\
		\left| \|\vecu^{\text{disc}}(\vecX^{(n_{\text{dof}})})\|_2 - \|\vecu^{\text{true}}(\vecX^{(n_{\text{dof}})})\|_2 \right|
	\end{pmatrix}.
\end{equation}

\noindent
For stress validation, we evaluate the von Mises equivalent stress at Gauss points as follows:
\begin{equation}
	\sigma_{\text{vM}} = \sqrt{(\sigma_{xx})^2 - \left(\sigma_{xx} \sigma_{yy}\right) + (\sigma_{yy})^2 + 3(\sigma_{xy})^2}
\end{equation}
The von Mises stress is projected from Gauss points onto  nodal coordinates, where we define the pointwise absolute von Mises stress error:
\begin{equation}\label{eq:point_svm_error}
	\vecvarepsilon_{\vecsigma_{\text{vM}}} =
	\begin{pmatrix}
		\left| \sigma_{\text{vM}}^{\text{disc}}\left(\vecX^{(1)}\right) - \sigma_{\text{vM}}^{\text{true}}\left(\vecX^{(1)}\right) \right| \\
		\vdots                                                                                                                             \\
		\left| \sigma_{\text{vM}}^{\text{disc}}\left(\vecX^{(n_{\text{dof}})}\right) - \sigma_{\text{vM}}^{\text{true}}\left(\vecX^{(n_{\text{dof}})}\right) \right|
	\end{pmatrix}
\end{equation}

\vspace{0.5em}
\noindent \textbf{EUCLID Discovery.} To initiate model discovery, a mesh is constructed around the sensor locations. This is necessary because the feature matrix \( \matA \) in \eqref{eq:feature_matrix} depends on deformation gradients computed within elements. The inverse problem is then solved using the EUCLID approach \eqref{eq:PaperC_euclid_final}. Throughout this study, we choose $\lambda_{\min} = 10^{-2}$, $\lambda_{\max} = 10^{10}$, and $N_\lambda = 1000$ \eqref{eq:lambda_min_max}. The threshold for $\text{RMSE}_{\lambda}$ is chosen as $\tau = 70$ and therefore the admissible interval for the regularization parameter is $\Lambda_{\mathrm{acc}} = [10^{-2},\,10^{4}]$ \eqref{eq:lambda_acc}.

\vspace{0.5em}
\noindent \textbf{statFEM-EUCLID Iterative Discovery.} The proposed iterative framework is initialized with a naive linear elastic \Rev{material model. It is used to compute forecasted displacements in the first iteration of the framework. While a more sophisticated constitutive model could be employed if expert knowledge were available, we deliberately select this simple, model-agnostic baseline to demonstrate the robustness of the discovery algorithm.} In both test cases, we consider
\begin{equation}
	E = 1.35\, \si{MPa}, \qquad \nu = 0.35.
	\label{eq:PaperC_LE_material_parameter}
\end{equation}
We then apply the statFEM-EUCLID framework outlined in \autoref{sec:euclid} and specified by \eqref{eq:PaperC_discovery_exntedned}.


\subsection{Neo--Hookean}\label{sec:NH}
We first consider a compressible Neo--Hookean material model as ground truth. The associated strain energy density function is given by:
\begin{equation}
	W^{\text{true}}(J_1, J_3) = A_{10} \left(J_1 - 3 \right) + \frac{K}{2} \left(J_3 - 1 \right)^2,
	\label{eq:PaperC_Neo--Hookean_strain_energy}
\end{equation}
with material parameters \( A_{10} = 0.5\, \si{MPa} \) and \( B_1 = \frac{K}{2} = 1.5\, \si{MPa} \). This corresponds to a small-strain shear modulus \( \mu = 2A_{10} = 1.0\, \si{MPa} \) and a Young's modulus \( E = 2.7\, \si{MPa} \).
\begin{table}[t]
	\hspace{-1cm}
	\begin{tabular}{lll|l|l|l|}
		\cline{4-6}
		                                                   &                                             &                                                         & Strain Energy Density Function                                                                                 & $\epsilon_{\vecu}$               & $\epsilon_{W}$                     \\ \hline\hline
		\multicolumn{1}{|l|}{$\sigma_{\vece}$}             & \multicolumn{1}{l|}{$\nsen$}                & \rule{0pt}{3.0ex}Truth                                  & $0.500\,(J_{1}-3) + 1.500\,{\left(J_{3}-1\right)}^2$                                                           & --                               & --                                 \\ \hline
		\multicolumn{1}{|l|}{}                             & \multicolumn{1}{l|}{}                       & \rule{0pt}{3.0ex}\cellcolor[HTML]{e6f5ff}EUCLID         & \cellcolor[HTML]{e6f5ff}--                                                                                     & \cellcolor[HTML]{e6f5ff}--       & \cellcolor[HTML]{e6f5ff}--         \\ \cline{3-6}
		\multicolumn{1}{|l|}{}                             & \multicolumn{1}{l|}{\multirow{-2.5}{*}{3}}  & \rule{0pt}{3.0ex}\cellcolor[HTML]{fff0e6}statFEM-EUCLID & \cellcolor[HTML]{fff0e6}$0.463\,(J_{1}-3) + 0.037\,(J_{2}-3) + 1.503\,{\left(J_{3}-1\right)}^2$                & \cellcolor[HTML]{fff0e6}$0.0427$ & \cellcolor[HTML]{fff0e6}$0.015801$ \\ \cline{2-6}
		\multicolumn{1}{|l|}{}                             & \multicolumn{1}{l|}{}                       & \rule{0pt}{3.0ex}\cellcolor[HTML]{e6f5ff}EUCLID         & \cellcolor[HTML]{e6f5ff}$0.427\,(J_{1}-3) + 2.002\,{\left(J_{2}-3\right)}^3 + 1.230\,{\left(J_{3}-1\right)}^2$ & \cellcolor[HTML]{e6f5ff}$0.0788$ & \cellcolor[HTML]{e6f5ff}$0.655804$ \\ \cline{3-6}
		\multicolumn{1}{|l|}{}                             & \multicolumn{1}{l|}{\multirow{-2.5}{*}{13}} & \rule{0pt}{3.0ex}\cellcolor[HTML]{fff0e6}statFEM-EUCLID & \cellcolor[HTML]{fff0e6}$0.483\,(J_{1}-3) + 1.449\,{\left(J_{3}-1\right)}^2$                                   & \cellcolor[HTML]{fff0e6}$0.0403$ & \cellcolor[HTML]{fff0e6}$0.001138$ \\ \cline{2-6}
		\multicolumn{1}{|l|}{}                             & \multicolumn{1}{l|}{}                       & \rule{0pt}{3.0ex}\cellcolor[HTML]{e6f5ff}EUCLID         & \cellcolor[HTML]{e6f5ff}$0.442\,(J_{1}-3) + 0.336\,{\left(J_{2}-3\right)}^2 + 1.438\,{\left(J_{3}-1\right)}^2$ & \cellcolor[HTML]{e6f5ff}$0.0134$ & \cellcolor[HTML]{e6f5ff}$0.259605$ \\ \cline{3-6}
		\multicolumn{1}{|l|}{\multirow{-6}{*}{$10^{-03}$}} & \multicolumn{1}{l|}{\multirow{-2.5}{*}{38}} & \rule{0pt}{3.0ex}\cellcolor[HTML]{fff0e6}statFEM-EUCLID & \cellcolor[HTML]{fff0e6}$0.484\,(J_{1}-3) + 1.452\,{\left(J_{3}-1\right)}^2$                                   & \cellcolor[HTML]{fff0e6}$0.0100$ & \cellcolor[HTML]{fff0e6}$0.001018$ \\ \hline
		\multicolumn{1}{|l|}{}                             & \multicolumn{1}{l|}{}                       & \rule{0pt}{3.0ex}\cellcolor[HTML]{e6f5ff}EUCLID         & \cellcolor[HTML]{e6f5ff}--                                                                                     & \cellcolor[HTML]{e6f5ff}--       & \cellcolor[HTML]{e6f5ff}--         \\ \cline{3-6}
		\multicolumn{1}{|l|}{}                             & \multicolumn{1}{l|}{\multirow{-2.5}{*}{3}}  & \rule{0pt}{3.0ex}\cellcolor[HTML]{fff0e6}statFEM-EUCLID & \cellcolor[HTML]{fff0e6}$0.496\,(J_{1}-3) + 1.488\,{\left(J_{3}-1\right)}^2$                                   & \cellcolor[HTML]{fff0e6}$0.0029$ & \cellcolor[HTML]{fff0e6}$0.000067$ \\ \cline{2-6}
		\multicolumn{1}{|l|}{}                             & \multicolumn{1}{l|}{}                       & \rule{0pt}{3.0ex}\cellcolor[HTML]{e6f5ff}EUCLID         & \cellcolor[HTML]{e6f5ff}$0.435\,(J_{1}-3) + 1.612\,{\left(J_{2}-3\right)}^3 + 1.239\,{\left(J_{3}-1\right)}^2$ & \cellcolor[HTML]{e6f5ff}$0.0806$ & \cellcolor[HTML]{e6f5ff}$0.446189$ \\ \cline{3-6}
		\multicolumn{1}{|l|}{}                             & \multicolumn{1}{l|}{\multirow{-2.5}{*}{13}} & \rule{0pt}{3.0ex}\cellcolor[HTML]{fff0e6}statFEM-EUCLID & \cellcolor[HTML]{fff0e6}$0.479\,(J_{1}-3) + 1.488\,{\left(J_{3}-1\right)}^2$                                   & \cellcolor[HTML]{fff0e6}$0.0018$ & \cellcolor[HTML]{fff0e6}$0.000002$ \\ \cline{2-6}
		\multicolumn{1}{|l|}{}                             & \multicolumn{1}{l|}{}                       & \rule{0pt}{3.0ex}\cellcolor[HTML]{e6f5ff}EUCLID         & \cellcolor[HTML]{e6f5ff}$0.464\,(J_{1}-3) + 0.714\,{\left(J_{2}-3\right)}^3 + 1.339\,{\left(J_{3}-1\right)}^2$ & \cellcolor[HTML]{e6f5ff}$0.0539$ & \cellcolor[HTML]{e6f5ff}$0.150306$ \\ \cline{3-6}
		\multicolumn{1}{|l|}{\multirow{-6}{*}{$10^{-04}$}} & \multicolumn{1}{l|}{\multirow{-2.5}{*}{38}} & \rule{0pt}{3.0ex}\cellcolor[HTML]{fff0e6}statFEM-EUCLID & \cellcolor[HTML]{fff0e6}$0.499\,(J_{1}-3) + 1.498\,{\left(J_{3}-1\right)}^2$                                   & \cellcolor[HTML]{fff0e6}$0.0014$ & \cellcolor[HTML]{fff0e6}$0.000001$ \\ \hline
	\end{tabular}
	\caption{Comparison of Neo--Hookean model as ground-truth with discovered constitutive model using the reconstructed full-fields and sparse measurements. The results are presented for different levels of measurement noise $\sigma_{\vece} = 10^{-03}, 10^{-04}$, number of sensors $ n_{\text{sen}} = 3, 13, 38$ and number of sensor readings $n_{\text{r}} = 1$ for $\epsilon_{\vecu}$ and $\epsilon_W$, see~\eqref{eq:PaperC_relative_rmse} and \eqref{eq:PaperC_relative_rmse_u}.}
	\label{tab:PaperC_discovered_models_NH}
\end{table}

\autoref{tab:PaperC_discovered_models_NH} presents a comparative summary between the ground-truth Neo--Hookean model and the constitutive models discovered using the EUCLID and statFEM-EUCLID frameworks. The results span a range of sensor noise levels and three sensor configurations. For each setting, we report the discovered strain energy density function, and the relative errors in strain energy density \( \epsilon_W \) \eqref{eq:PaperC_relative_rmse} and displacement  \( \epsilon_{\vecu} \) \eqref{eq:PaperC_relative_rmse_u}.

\begin{figure}[htp!]
	\centering
	\subfloat[Iteration $1$ to $10$]{
		\centering
		\includegraphics{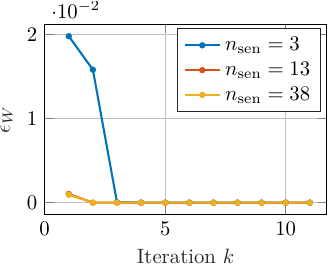}
		\label{subfig:convergence_eW_1_10_e4}}
	\hspace*{1cm}
	\subfloat[Iteration $3$ to $10$]{
		\centering
		\includegraphics{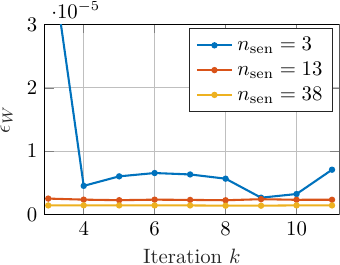}
		\label{subfig:convergence_eW_3_10_e4}}
	\caption{Convergence of the relative strain energy reconstruction error \( \epsilon_W \) over model discovery iterations for $\sigma_{\vece}=10^{-4}$. (a) Full iteration history. (b) Zoomed view for iterations \( k \geq 3 \) to highlight convergence stability.}
	\label{fig:convergence_eW}
\end{figure}

\begin{figure}[H]
	\centering
	\subfloat[$\nsen =13$, $\sigma_{\vece} = 10^{-03}$]{
		\centering
		\includegraphics{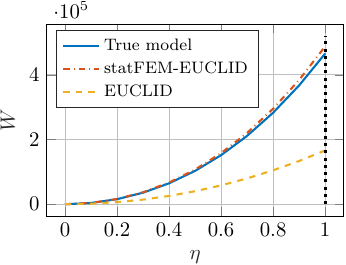}
		\label{subfig:plateWithHole_energy_vs_f_nsen13_e3}}
	\hspace*{0.5cm}
	\subfloat[$\nsen =38$, $\sigma_{\vece} = 10^{-03}$]{
		\centering
		\includegraphics{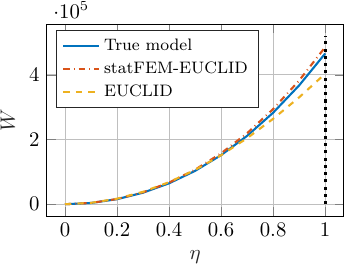}
		\label{subfig:plateWithHole_energy_vs_f_nsen38_e3}} \\
	\subfloat[$\nsen =13$, $\sigma_{\vece} = 10^{-04}$]{
		\centering
		\includegraphics{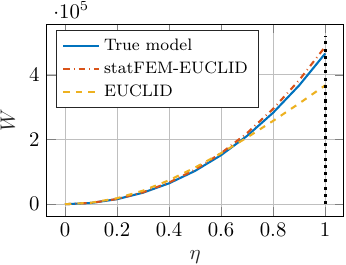}
		\label{subfig:plateWithHole_energy_vs_f_nsen13_e4}}
	\hspace*{0.5cm}
	\subfloat[$\nsen =38$, $\sigma_{\vece} = 10^{-04}$]{
		\centering
		\includegraphics{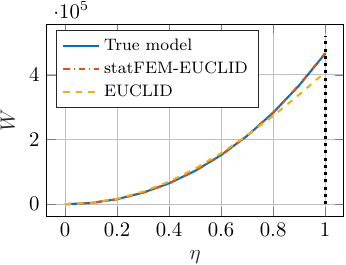}
		\label{subfig:plateWithHole_energy_vs_f_nsen38_e4}}
	\caption{Comparison of the strain energy function $W$ against the nondimensional load scale $\eta\in[0,1]$ between the true Neo--Hookean model, EUCLID, and statFEM-EUCLID for different sensor configurations and noise levels. The model discovery is performed on $\eta=1$. Results are depicted for the element marked in \autoref{fig:plateWithHole_mesh}.}
	\label{fig:plateWithHole_energy_vs_f}
\end{figure}

\noindent
The convergence plots in \autoref{fig:convergence_eW} demonstrate that the statFEM-EUCLID framework reliably identifies the correct strain energy model within a few iterations. As seen in \autoref{subfig:convergence_eW_1_10_e4}, the reconstruction error \( \epsilon_W \) drops significantly in the first 2--3 iterations, even with only \(\nsen = 3\) sensors. \autoref{subfig:convergence_eW_3_10_e4} shows that beyond iteration \(k = 3\), all curves flatten, indicating convergence. At a noise level of $\sigma_{\vece}=10^{-4}$, with both \(\nsen = 13\) and \(38\), the final error remains below \(10^{-5}\), which confirms accurate and stable model discovery.
\begin{figure}[!t]
	\centering
	\hspace*{-0.5cm}
	\subfloat[$\nsen =13$, $\sigma_{\vece} = 10^{-03}$]{
		\centering
		\hspace*{-0.7cm}
		\includegraphics{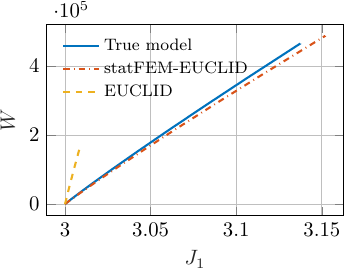}
		\includegraphics{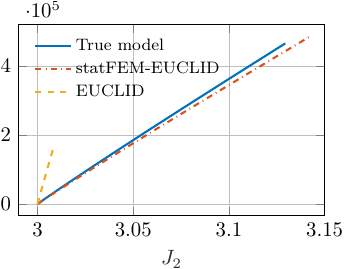}
		\hspace*{-0.4cm}
		\includegraphics{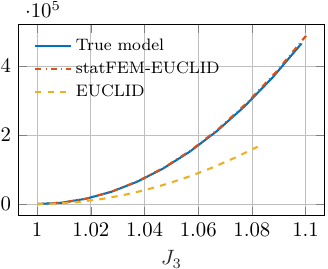}
		\label{subfig:plateWithHole_energy_vs_J_nsen13_e3}} \\
	\centering
	\hspace*{-0.5cm}
	\subfloat[$\nsen =38$, $\sigma_{\vece} = 10^{-03}$]{
		\centering
		\hspace*{-0.7cm}
		\includegraphics{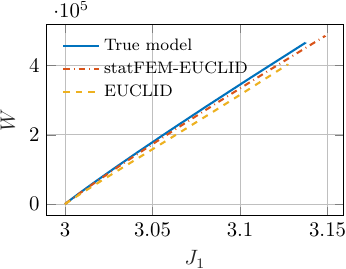}
		\includegraphics{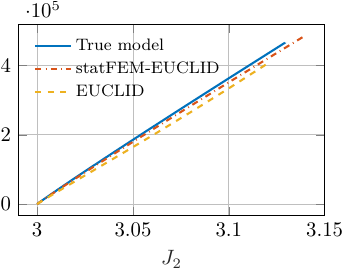}
		\hspace*{-0.4cm}
		\includegraphics{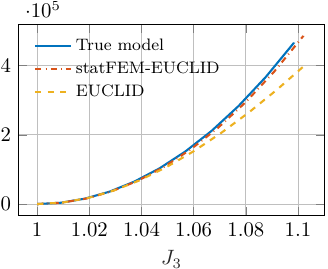}
		\label{subfig:plateWithHole_energy_vs_J_nsen38_e3}}
	\caption{Comparison of the strain energy function $W$ against the modified invariants $J_1$, $J_2$, and $J_3$ between the true Neo--Hookean model, EUCLID, and statFEM-EUCLID for different sensor configurations and noise levels $\sigma_{\vece} = 10^{-3}$. The invariants are controlled via the load scale $\eta\in[0,1]$, c.f.~\autoref{fig:plateWithHole_energy_vs_f}. Results are depicted for the element marked in \autoref{fig:plateWithHole_mesh}.}
	\label{fig:plateWithHole_energy_vs_J_e3}
\end{figure}

\begin{figure}[!t]
	\centering
	\hspace*{-0.5cm}
	\subfloat[$\nsen =13$, $\sigma_{\vece} = 10^{-04}$]{
		\centering
		\hspace*{-0.7cm}
		\includegraphics{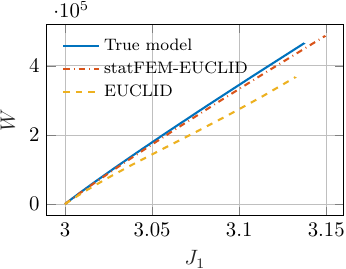}
		\includegraphics{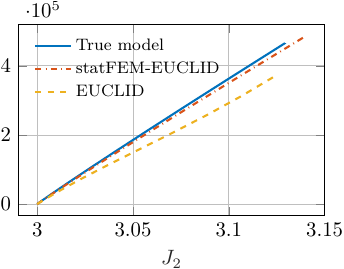}
		\hspace*{-0.4cm}
		\includegraphics{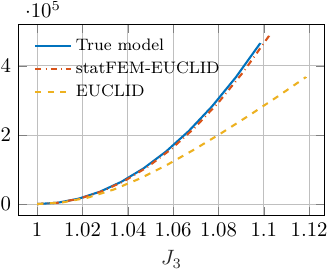}
		\label{subfig:plateWithHole_energy_vs_J_nsen13_e4}} \\
	\centering
	\hspace*{-0.5cm}
	\subfloat[$\nsen =38$, $\sigma_{\vece} = 10^{-04}$]{
		\centering
		\hspace*{-0.7cm}
		\includegraphics{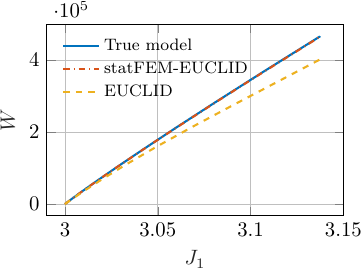}
		\hspace*{-0.4cm}
		\includegraphics{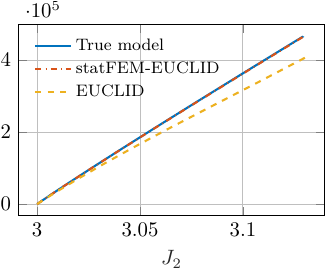}
		\hspace*{-0.1cm}
		\includegraphics{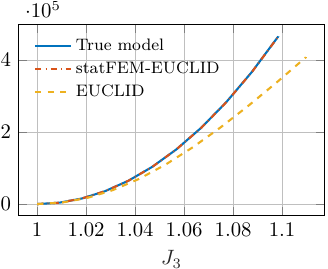}
		\label{subfig:plateWithHole_energy_vs_J_nsen38_e4}}
	\caption{Comparison of the strain energy function $W$ against the invariants $J_1$, $J_2$, and $J_3$ between the true Neo--Hookean model, EUCLID, and statFEM-EUCLID for different sensor configurations and noise levels $\sigma_{\vece} = 10^{-4}$. The invariants are controlled via the load scale $\eta\in[0,1]$, c.f.~\autoref{fig:plateWithHole_energy_vs_f}. Results are depicted for the element marked in \autoref{fig:plateWithHole_mesh}.}
	\label{fig:plateWithHole_energy_vs_J_e4}
\end{figure}

\autoref{fig:plateWithHole_energy_vs_f} compares the strain energy function $W$ predicted by the discovered models against the Neo--Hookean ground truth as a function of the nondimensional load scale $\eta$ for a selected element marked in \autoref{fig:plateWithHole_mesh}. The results demonstrate that the proposed statFEM-EUCLID framework achieves a significantly closer match to the true model than EUCLID, particularly in the presence of noise. For the case of $\nsen = 13$ sensors with $\sigma_{\vece} = 10^{-3}  ($\autoref{subfig:plateWithHole_energy_vs_f_nsen13_e3}), EUCLID underestimates the strain energy at larger load scales, while statFEM-EUCLID closely tracks the true response across the full range of $\eta$. Increasing the number of sensors to $\nsen = 38$ improves the results for both methods. Still, statFEM-EUCLID consistently outperforms EUCLID, especially near $\eta = 1$, where accurate recovery of nonlinear material behavior is most critical. At lower noise levels ($\sigma_{\vece} = 10^{-4}$), both frameworks benefit from improved data quality. Yet, statFEM-EUCLID continues to deliver the most faithful reconstruction, essentially overlapping with the ground truth for $\nsen = 38$, see~\autoref{subfig:plateWithHole_energy_vs_f_nsen38_e4}.

\begin{figure}[!b]
	\centering
	\subfloat[$\vecu^{\text{true}}$]{
		\centering
		\includegraphics{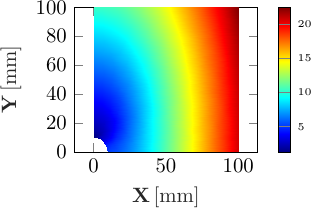}
		\label{subfig:plateWithHole_disp_true_contour}}
	\subfloat[$\vecu^{\text{disc}}$]{
		\centering
		\includegraphics{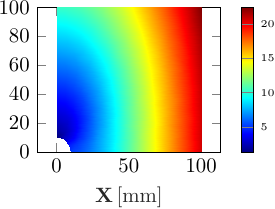}
		\label{subfig:plateWithHole_dispError_nsen15}}
	\subfloat[$\vecvarepsilon_{\vecu}$]{
		\centering
		\includegraphics{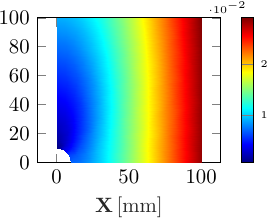}
		\label{subfig:plateWithHole_dispError_nsen38}}
	\caption{True and inferred displacement with Neo--Hookean model.}
	\label{fig:plateWithHole_contour_displacement}
\end{figure}

The comparisons in \autoref{fig:plateWithHole_energy_vs_J_e3} and \autoref{fig:plateWithHole_energy_vs_J_e4} show the strain energy function $W$ in the element marked in \autoref{fig:plateWithHole_mesh} as a function of the modified invariants $J_1(\eta)$, $J_2(\eta)$, and $J_3(\eta)$ for different sensor configurations and noise levels. The invariants are controlled via the load scale $\eta$. Across all cases, the statFEM-EUCLID framework provides a significantly closer approximation to the true Neo--Hookean response than EUCLID. In particular, deviations are most pronounced in the volumetric response ($J_3$), where EUCLID systematically underestimates the strain energy. At the same time, statFEM-EUCLID nearly overlaps with the ground truth, even under higher noise ($\sigma_{\vece} = 10^{-3}$). For the isochoric invariants $J_1$ and $J_2$, statFEM-EUCLID again achieves strong agreement, with only minor offsets at large invariant values. Increasing the number of sensors from $n_{\text{sen}} = 13$ to $n_{\text{sen}} = 38$ further improves accuracy for both methods, though the advantage of statFEM-EUCLID over EUCLID remains evident. An alternative comparison of the strain energy density function $W$ against the individual components of the deformation gradient $\matF$ in Euclidean space is provided in \autoref{sec:PaperC_additional_results}.

\begin{figure}[!ht]
	\centering
	\subfloat[$\vecsigma^{\text{true}}_{\text{vM}}$]{
		\centering
		\includegraphics{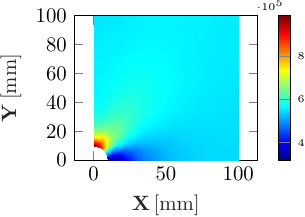}
		\label{subfig:plateWithHole_vonMises_true_contour}}
	\subfloat[$\vecsigma^{\text{disc}}_{\text{vM}}$]{
		\centering
		\includegraphics{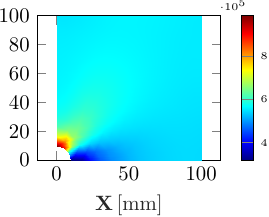}
		\label{subfig:plateWithHole_vonMises_discovered_contour}}
	\subfloat[$\vecvarepsilon_{\vecsigma_{\text{vM}}}$]{
		\centering
		\includegraphics{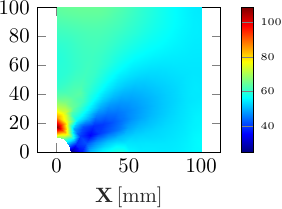}
		\label{subfig:plateWithHole_vonMises_error_nsen38}}
	\caption{True and inferred Cauchy von Mises stress with Neo--Hookean model.}
	\label{fig:plateWithHole_contour_vonMises}
\end{figure}

\autoref{fig:plateWithHole_contour_displacement} and \autoref{fig:plateWithHole_contour_vonMises} demonstrate the fidelity of the discovered Neo--Hookean model by comparing the associated displacement and von Mises stress fields against the true solution, with error metrics defined in \eqref{eq:point_disp_error} and \eqref{eq:point_svm_error}. The recovered displacement magnitude (\autoref{subfig:plateWithHole_dispError_nsen15}) closely matches the ground truth (\autoref{subfig:plateWithHole_disp_true_contour}), with the error field (\autoref{subfig:plateWithHole_dispError_nsen38}) remaining below \(2 \times 10^{-2} \, \si{mm}\) across the domain. Similarly, the inferred von Mises stress field (\autoref{subfig:plateWithHole_vonMises_discovered_contour}) captures the localized stress concentrations with good agreement, and the corresponding error (\autoref{subfig:plateWithHole_vonMises_error_nsen38}) is mostly confined near boundary features.

\subsection{Mooney--Rivlin}\label{sec:MR}

We now consider a second synthetic benchmark using a Mooney--Rivlin model as the ground truth. The strain energy density function is defined as
\begin{equation}
	W^{\text{true}}(J_1, J_2, J_3) = A_{10} \left(J_1 - 3 \right)+ A_{01} \, (J_2 - 3) + \frac{K}{2} \left(J_3 - 1 \right)^2,
	\label{eq:PaperC_mooney_rivlin_strain_energy}
\end{equation}
which involves two isochoric and one volumetric parameter. Here, $A_{10} = 0.3\, \si{MPa}$, $A_{01} = 0.2\, \si{MPa}$, and \( B_1 = \frac{K}{2} = 1.5\, \si{MPa} \). Synthetic displacement measurements are generated using the same procedure as described in~\autoref{sec:PaperC_NumExamples}, with added Gaussian noise. We examine the performance of both EUCLID and statFEM-EUCLID across different sensor configurations and noise levels.

\autoref{tab:PaperC_discovered_models_MR} summarizes the discovered strain energy models, and the relative errors in strain energy density \( \epsilon_W \) \eqref{eq:PaperC_relative_rmse} and displacement  \( \epsilon_{\vecu} \) \eqref{eq:PaperC_relative_rmse_u}. As in the Neo--Hookean case, statFEM-EUCLID consistently outperforms EUCLID, achieving significantly lower errors across all configurations. In particular, the statFEM-EUCLID framework can accurately recover both isochoric terms, even under sparse sensing conditions and in the presence of higher noise. With just three sensors and low noise (\(\sigma_{\vece} = 10^{-4}\)), it produces a model with displacement and energy errors of \(\epsilon_{\vecu} = 0.0180\) and \(\epsilon_W = 0.0762\), respectively. As the number of sensors increases or the noise decreases, both errors approach near-zero levels. For example, at \( \nsen = 38 \) and \( \sigma_{\vece} = 10^{-4} \), statFEM-EUCLID achieves errors on the order of \( \epsilon_W \leq 4 \times 10^{-5} \).

The Mooney--Rivlin case presents a more challenging scenario, since the true strain energy density involves more invariants. Nevertheless, the statFEM-EUCLID framework again demonstrates strong performance. When $\nsen =13$, the discovered models approximate the relevant terms of the Mooney--Rivlin form. With the increasing availability of sensors ($\nsen = 38$) and reduced noise, the framework progressively refines its estimates, converging toward the correct functional form and accurate coefficients. In contrast, EUCLID-only discovery struggles with noisy or sparse input data, often misidentifying higher-order terms.

As the results reported in \autoref{tab:PaperC_discovered_models_NH} and \autoref{tab:PaperC_discovered_models_MR} demonstrate a similar trend for both test cases, we refrain from repeating the exhaustive evaluation reported for the Neo-Hookean case.

\begin{table}[htp!]
	\hspace{-1cm}
	\begin{tabular}{lll|l|l|l|}
		\cline{4-6}
		                                                   &                                             &                                                         & Strain Energy Density Function                                                                                 & $\epsilon_{\vecu}$               & $\epsilon_{W}$                     \\ \hline\hline
		\multicolumn{1}{|l|}{$\sigma_{\vece}$}             & \multicolumn{1}{l|}{$\nsen$}                & \rule{0pt}{3.0ex}Truth                                  & $0.3\,(J_{1}-3) + 0.2\,(J_{2}-3) + 1.5\,{\left(J_{3}-1\right)}^2$                                              & --                               & --                                 \\ \hline
		\multicolumn{1}{|l|}{}                             & \multicolumn{1}{l|}{}                       & \rule{0pt}{3.0ex}\cellcolor[HTML]{e6f5ff}EUCLID         & \cellcolor[HTML]{e6f5ff}--                                                                                     & \cellcolor[HTML]{e6f5ff}--       & \cellcolor[HTML]{e6f5ff}--         \\ \cline{3-6}
		\multicolumn{1}{|l|}{}                             & \multicolumn{1}{l|}{\multirow{-2.5}{*}{3}}  & \rule{0pt}{3.0ex}\cellcolor[HTML]{fff0e6}statFEM-EUCLID & \cellcolor[HTML]{fff0e6}$0.407\,(J_{1}-3) + 1.220\,{\left(J_{3}-1\right)}^2$                                   & \cellcolor[HTML]{fff0e6}$0.2556$ & \cellcolor[HTML]{fff0e6}$0.084210$ \\ \cline{2-6}
		\multicolumn{1}{|l|}{}                             & \multicolumn{1}{l|}{}                       & \rule{0pt}{3.0ex}\cellcolor[HTML]{e6f5ff}EUCLID         & \cellcolor[HTML]{e6f5ff}$0.425\,(J_{1}-3) + 1.368\,{\left(J_{2}-3\right)}^3 + 1.178\,{\left(J_{3}-1\right)}^2$ & \cellcolor[HTML]{e6f5ff}$0.0864$ & \cellcolor[HTML]{e6f5ff}$0.057923$ \\ \cline{3-6}
		\multicolumn{1}{|l|}{}                             & \multicolumn{1}{l|}{\multirow{-2.5}{*}{13}} & \rule{0pt}{3.0ex}\cellcolor[HTML]{fff0e6}statFEM-EUCLID & \cellcolor[HTML]{fff0e6}$0.322\,(J_{1}-3) + 0.175\,(J_{2}-3) + 1.491\,{\left(J_{3}-1\right)}^2$                & \cellcolor[HTML]{fff0e6}$0.0049$ & \cellcolor[HTML]{fff0e6}$0.001850$ \\ \cline{2-6}
		\multicolumn{1}{|l|}{}                             & \multicolumn{1}{l|}{}                       & \rule{0pt}{3.0ex}\cellcolor[HTML]{e6f5ff}EUCLID         & \cellcolor[HTML]{e6f5ff}$0.434\,(J_{1}-3) + 0.307\,{\left(J_{2}-3\right)}^2 + 1.381\,{\left(J_{3}-1\right)}^2$ & \cellcolor[HTML]{e6f5ff}$0.0132$ & \cellcolor[HTML]{e6f5ff}$0.082018$ \\ \cline{3-6}
		\multicolumn{1}{|l|}{\multirow{-6}{*}{$10^{-03}$}} & \multicolumn{1}{l|}{\multirow{-2.5}{*}{38}} & \rule{0pt}{3.0ex}\cellcolor[HTML]{fff0e6}statFEM-EUCLID & \cellcolor[HTML]{fff0e6}$0.328\,(J_{1}-3) + 0.169\,(J_{2}-3) + 1.490\,{\left(J_{3}-1\right)}^2$                & \cellcolor[HTML]{fff0e6}$0.0048$ & \cellcolor[HTML]{fff0e6}$0.001140$ \\ \hline
		\multicolumn{1}{|l|}{}                             & \multicolumn{1}{l|}{}                       & \rule{0pt}{3.0ex}\cellcolor[HTML]{e6f5ff}EUCLID         & \cellcolor[HTML]{e6f5ff}--                                                                                     & \cellcolor[HTML]{e6f5ff}--       & \cellcolor[HTML]{e6f5ff}--         \\ \cline{3-6}
		\multicolumn{1}{|l|}{}                             & \multicolumn{1}{l|}{\multirow{-2.5}{*}{3}}  & \rule{0pt}{3.0ex}\cellcolor[HTML]{fff0e6}statFEM-EUCLID & \cellcolor[HTML]{fff0e6}$0.481\,(J_{1}-3) + 1.444\,{\left(J_{3}-1\right)}^2$                                   & \cellcolor[HTML]{fff0e6}$0.0180$ & \cellcolor[HTML]{fff0e6}$0.076238$ \\ \cline{2-6}
		\multicolumn{1}{|l|}{}                             & \multicolumn{1}{l|}{}                       & \rule{0pt}{3.0ex}\cellcolor[HTML]{e6f5ff}EUCLID         & \cellcolor[HTML]{e6f5ff}$0.419\,(J_{1}-3) + 1.687\,{\left(J_{2}-3\right)}^3 + 1.168\,{\left(J_{3}-1\right)}^2$ & \cellcolor[HTML]{e6f5ff}$0.0847$ & \cellcolor[HTML]{e6f5ff}$0.066547$ \\ \cline{3-6}
		\multicolumn{1}{|l|}{}                             & \multicolumn{1}{l|}{\multirow{-2.5}{*}{13}} & \rule{0pt}{3.0ex}\cellcolor[HTML]{fff0e6}statFEM-EUCLID & \cellcolor[HTML]{fff0e6}$0.287\,(J_{1}-3) + 0.212\,(J_{2}-3) + 1.496\,{\left(J_{3}-1\right)}^2$                & \cellcolor[HTML]{fff0e6}$0.0041$ & \cellcolor[HTML]{fff0e6}$0.000323$ \\ \cline{2-6}
		\multicolumn{1}{|l|}{}                             & \multicolumn{1}{l|}{}                       & \rule{0pt}{3.0ex}\cellcolor[HTML]{e6f5ff}EUCLID         & \cellcolor[HTML]{e6f5ff}$0.456\,(J_{1}-3) + 0.628\,{\left(J_{2}-3\right)}^3 + 1.291\,{\left(J_{3}-1\right)}^2$ & \cellcolor[HTML]{e6f5ff}$0.0514$ & \cellcolor[HTML]{e6f5ff}$0.078472$ \\ \cline{3-6}
		\multicolumn{1}{|l|}{\multirow{-6}{*}{$10^{-04}$}} & \multicolumn{1}{l|}{\multirow{-2.5}{*}{38}} & \rule{0pt}{3.0ex}\cellcolor[HTML]{fff0e6}statFEM-EUCLID & \cellcolor[HTML]{fff0e6}$0.302\,(J_{1}-3) + 0.196\,(J_{2}-3) + 1.494\,{\left(J_{3}-1\right)}^2$                & \cellcolor[HTML]{fff0e6}$0.0040$ & \cellcolor[HTML]{fff0e6}$0.000038$ \\ \hline
	\end{tabular}
	\caption{Comparison of Mooney--Rivlin model as ground-truth with discovered constitutive model using the reconstructed full-fields and sparse measurements. The results are presented for different levels of measurement noise $\sigma_{\vece} = 10^{-03}, 10^{-04}$, number of sensors $ n_{\text{sen}} = 3, 13, 38$ and number of sensor readings $n_{\text{r}} = 1$ for $\epsilon_{\vecu}$ and $\epsilon_W$, See~\eqref{eq:PaperC_relative_rmse} and \eqref{eq:PaperC_relative_rmse_u}.}
	\label{tab:PaperC_discovered_models_MR}
\end{table}

\section{Conclusion and Outlook}\label{sec:PaperC_Conclusion_Outlook}

In this work, we have presented a framework for the unsupervised discovery of constitutive models from sparse and noisy displacement data. By integrating the statistical Finite Element Method (statFEM) with the Virtual Fields Method (VFM) and EUCLID, we address a central challenge in computational mechanics: how to reliably discover constitutive models when experimental data are corrupted by noise and available only at a limited number of sensor locations.

The key innovation lies in the iterative coupling of Bayesian state assimilation and sparse regression-based constitutive model discovery. StatFEM assimilates noisy and gappy measurements into a probabilistic full-field displacement estimate, which serves as a robust surrogate for the true response. This assimilated field then provides the input for EUCLID, enabling accurate and parsimonious identification of constitutive models. By enforcing equilibrium in the weak form and promoting sparsity through $\ell_1$-regularization, the method is capable of isolating the minimal set of features necessary to explain the observed data, while maintaining physical interpretability.

Our numerical studies with Neo--Hookean and Mooney--Rivlin materials demonstrate several essential properties of the proposed framework. First, statFEM-EUCLID achieves high-fidelity recovery of the ground truth model, even under measurement noise and sparse sensing, whereas EUCLID alone often fails to do so. In particular, near-perfect reconstruction was achieved in the Neo--Hookean case with dense sensing and low noise. Moreover, the more complex Mooney--Rivlin case demonstrated that the method can also successfully handle constitutive models involving multiple invariants. Second, the approach remains applicable in recovering meaningful material features with as few as three sensors. Finally, the Bayesian assimilation step proves essential for mitigating the effects of noisy data, which enables more stable parameter identification and discovery compared to VFM- or EUCLID-only approaches.

\section*{CRediT authorship contribution statement}
\textbf{Vahab Narouie (VN):} Conceptualization, Methodology, Formal Analysis, Software, Writing - Original Draft, Visualization.
\textbf{Jorge-Humberto Urrea-Quintero (JHUQ):} Conceptualization, Methodology, Formal Analysis, Writing - Review \& Editing.
\textbf{Fehmi Cirak (FC):} Conceptualization, Methodology, Writing - Review \& Editing.
\textbf{Henning Wessels (HW):} Conceptualization, Methodology, Writing - Original Draft, Writing - Review \& Editing, Funding Acquisition.

\section*{Declaration of competing interest}
The authors declare that they have no known competing financial interests or personal relationships that could have appeared to influence the work reported in this paper.
\section*{Acknowledgement}\label{sec:PaperC_Ackn}

The authors would like to thank Ulrich Römer and David Anton for their fruitful discussions. The support of the German Research Foundation is gratefully acknowledged in the following projects:
\begin{itemize}
	\item DFG 501798687 (HW): \textit{Monitoring data-driven life cycle management with AR based on adaptive, AI-supported corrosion prediction for reinforced concrete structures under combined impacts}. Subproject of SPP 2388: \textit{Hundred plus - Extending the Lifetime of Complex Engineering Structures through Intelligent Digitalization}.
	\item BMUV 02E12244C (HW): \textit{Verbundprojekt: KI-unterstützte Stoffmodellierung am Beispiel von Bentonit (KI-Stoff), Teilprojekt C}.
\end{itemize}

\section*{Data availability}
Upon acceptance of this manuscript, the source code supporting this study will be made available on GitHub and Zenodo \cite{knauf_narouie_2025_17249520}.

\begin{appendices}
	\renewcommand{\thetable}{\arabic{table}}

	\section{Model Library}\label{sec:PaperC_basis_functions}
	\begin{table}[ht]
		\centering
		\begin{tabular}{l|ll|ll|ll|ll|ll|l|}
			\cline{2-12}
			                                           & \multicolumn{2}{l|}{$N_{\text{MR}} = 1$} & \multicolumn{2}{l|}{$N_{\text{MR}} = 2$} & \multicolumn{2}{l|}{$N_{\text{MR}} = 3$} & \multicolumn{2}{l|}{$N_{\text{MR}} = 4$} & \multicolumn{2}{l|}{$N_{\text{MR}} = 5$} & \multicolumn{1}{c|}{\multirow{2}{*}{Features}}                                                                                                                                                                \\ \cline{2-11}
			                                           & \multicolumn{1}{l|}{$A_{ij}$}            & $\phi_i$                                 & \multicolumn{1}{l|}{$A_{ij}$}            & $\phi_i$                                 & \multicolumn{1}{l|}{$A_{ij}$}            & $\phi_i$                                       & \multicolumn{1}{l|}{$A_{ij}$}            & $\phi_i$                                 & \multicolumn{1}{l|}{$A_{ij}$} & $\phi_i$    &                          \\ \cline{2-12}
			                                           & \multicolumn{1}{l|}{$A_{10}$}            & $\phi_1$                                 & \multicolumn{1}{l|}{$A_{10}$}            & $\phi_1$                                 & \multicolumn{1}{l|}{$A_{10}$}            & $\phi_1$                                       & \multicolumn{1}{l|}{$A_{10}$}            & $\phi_1$                                 & \multicolumn{1}{l|}{$A_{10}$} & $\phi_1$    & $(J_1 - 3)$              \\
			                                           & \multicolumn{1}{l|}{$A_{01}$}            & $\phi_2$                                 & \multicolumn{1}{l|}{$A_{01}$}            & $\phi_2$                                 & \multicolumn{1}{l|}{$A_{01}$}            & $\phi_2$                                       & \multicolumn{1}{l|}{$A_{01}$}            & $\phi_2$                                 & \multicolumn{1}{l|}{$A_{01}$} & $\phi_2$    & $(J_2 - 3)$              \\
			                                           & \multicolumn{1}{c|}{\multirow{1}{*}{--}} & \multicolumn{1}{c|}{\multirow{1}{*}{--}} & \multicolumn{1}{c|}{$A_{20}$}            & $\phi_3$                                 & \multicolumn{1}{c|}{$A_{20}$}            & $\phi_3$                                       & \multicolumn{1}{c|}{$A_{20}$}            & $\phi_3$                                 & \multicolumn{1}{c|}{$A_{20}$} & $\phi_3$    & $(J_1 - 3)^2$            \\
			                                           & \multicolumn{1}{c|}{\multirow{1}{*}{--}} & \multicolumn{1}{c|}{\multirow{1}{*}{--}} & \multicolumn{1}{c|}{$A_{11}$}            & $\phi_4$                                 & \multicolumn{1}{c|}{$A_{11}$}            & $\phi_4$                                       & \multicolumn{1}{c|}{$A_{11}$}            & $\phi_4$                                 & \multicolumn{1}{c|}{$A_{11}$} & $\phi_4$    & $(J_1 - 3)(J_2 - 3)$     \\
			                                           & \multicolumn{1}{c|}{\multirow{1}{*}{--}} & \multicolumn{1}{c|}{\multirow{1}{*}{--}} & \multicolumn{1}{c|}{$A_{02}$}            & $\phi_5$                                 & \multicolumn{1}{c|}{$A_{02}$}            & $\phi_5$                                       & \multicolumn{1}{c|}{$A_{02}$}            & $\phi_5$                                 & \multicolumn{1}{c|}{$A_{02}$} & $\phi_5$    & $(J_2 - 3)^2$            \\
			                                           & \multicolumn{1}{c|}{\multirow{1}{*}{--}} & \multicolumn{1}{c|}{\multirow{1}{*}{--}} & \multicolumn{1}{c|}{\multirow{1}{*}{--}} & \multicolumn{1}{c|}{\multirow{1}{*}{--}} & \multicolumn{1}{c|}{$A_{30}$}            & $\phi_6$                                       & \multicolumn{1}{c|}{$A_{30}$}            & $\phi_6$                                 & \multicolumn{1}{c|}{$A_{30}$} & $\phi_6$    & $(J_1 - 3)^3$            \\
			                                           & \multicolumn{1}{c|}{\multirow{1}{*}{--}} & \multicolumn{1}{c|}{\multirow{1}{*}{--}} & \multicolumn{1}{c|}{\multirow{1}{*}{--}} & \multicolumn{1}{c|}{\multirow{1}{*}{--}} & \multicolumn{1}{c|}{$A_{21}$}            & $\phi_7$                                       & \multicolumn{1}{c|}{$A_{21}$}            & $\phi_7$                                 & \multicolumn{1}{c|}{$A_{21}$} & $\phi_7$    & $(J_1 - 3)^2(J_2 - 3)$   \\
			                                           & \multicolumn{1}{c|}{\multirow{1}{*}{--}} & \multicolumn{1}{c|}{\multirow{1}{*}{--}} & \multicolumn{1}{c|}{\multirow{1}{*}{--}} & \multicolumn{1}{c|}{\multirow{1}{*}{--}} & \multicolumn{1}{c|}{$A_{12}$}            & $\phi_8$                                       & \multicolumn{1}{c|}{$A_{12}$}            & $\phi_8$                                 & \multicolumn{1}{c|}{$A_{12}$} & $\phi_8$    & $(J_1 - 3)(J_2 - 3)^2$   \\
			                                           & \multicolumn{1}{c|}{\multirow{1}{*}{--}} & \multicolumn{1}{c|}{\multirow{1}{*}{--}} & \multicolumn{1}{c|}{\multirow{1}{*}{--}} & \multicolumn{1}{c|}{\multirow{1}{*}{--}} & \multicolumn{1}{c|}{$A_{03}$}            & $\phi_9$                                       & \multicolumn{1}{c|}{$A_{03}$}            & $\phi_9$                                 & \multicolumn{1}{c|}{$A_{03}$} & $\phi_9$    & $(J_2 - 3)^3$            \\
			                                           & \multicolumn{1}{c|}{\multirow{1}{*}{--}} & \multicolumn{1}{c|}{\multirow{1}{*}{--}} & \multicolumn{1}{c|}{\multirow{1}{*}{--}} & \multicolumn{1}{c|}{\multirow{1}{*}{--}} & \multicolumn{1}{c|}{\multirow{1}{*}{--}} & \multicolumn{1}{c|}{\multirow{1}{*}{--}}       & \multicolumn{1}{l|}{$A_{40}$}            & $\phi_{10}$                              & \multicolumn{1}{l|}{$A_{40}$} & $\phi_{10}$ & $(J_1 - 3)^4$            \\
			                                           & \multicolumn{1}{c|}{\multirow{1}{*}{--}} & \multicolumn{1}{c|}{\multirow{1}{*}{--}} & \multicolumn{1}{c|}{\multirow{1}{*}{--}} & \multicolumn{1}{c|}{\multirow{1}{*}{--}} & \multicolumn{1}{c|}{\multirow{1}{*}{--}} & \multicolumn{1}{c|}{\multirow{1}{*}{--}}       & \multicolumn{1}{l|}{$A_{31}$}            & $\phi_{11}$                              & \multicolumn{1}{l|}{$A_{31}$} & $\phi_{11}$ & $(J_1 - 3)^3(J_2 - 3)$   \\
			                                           & \multicolumn{1}{c|}{\multirow{1}{*}{--}} & \multicolumn{1}{c|}{\multirow{1}{*}{--}} & \multicolumn{1}{c|}{\multirow{1}{*}{--}} & \multicolumn{1}{c|}{\multirow{1}{*}{--}} & \multicolumn{1}{c|}{\multirow{1}{*}{--}} & \multicolumn{1}{c|}{\multirow{1}{*}{--}}       & \multicolumn{1}{l|}{$A_{22}$}            & $\phi_{12}$                              & \multicolumn{1}{l|}{$A_{22}$} & $\phi_{12}$ & $(J_1 - 3)^2(J_2 - 3)^2$ \\
			                                           & \multicolumn{1}{c|}{\multirow{1}{*}{--}} & \multicolumn{1}{c|}{\multirow{1}{*}{--}} & \multicolumn{1}{c|}{\multirow{1}{*}{--}} & \multicolumn{1}{c|}{\multirow{1}{*}{--}} & \multicolumn{1}{c|}{\multirow{1}{*}{--}} & \multicolumn{1}{c|}{\multirow{1}{*}{--}}       & \multicolumn{1}{l|}{$A_{13}$}            & $\phi_{13}$                              & \multicolumn{1}{l|}{$A_{13}$} & $\phi_{13}$ & $(J_1 - 3)(J_2 - 3)^3$   \\
			                                           & \multicolumn{1}{c|}{\multirow{1}{*}{--}} & \multicolumn{1}{c|}{\multirow{1}{*}{--}} & \multicolumn{1}{c|}{\multirow{1}{*}{--}} & \multicolumn{1}{c|}{\multirow{1}{*}{--}} & \multicolumn{1}{c|}{\multirow{1}{*}{--}} & \multicolumn{1}{c|}{\multirow{1}{*}{--}}       & \multicolumn{1}{l|}{$A_{04}$}            & $\phi_{14}$                              & \multicolumn{1}{l|}{$A_{04}$} & $\phi_{14}$ & $(J_2 - 3)^4$            \\
			                                           & \multicolumn{1}{c|}{\multirow{1}{*}{--}} & \multicolumn{1}{c|}{\multirow{1}{*}{--}} & \multicolumn{1}{c|}{\multirow{1}{*}{--}} & \multicolumn{1}{c|}{\multirow{1}{*}{--}} & \multicolumn{1}{c|}{\multirow{1}{*}{--}} & \multicolumn{1}{c|}{\multirow{1}{*}{--}}       & \multicolumn{1}{c|}{\multirow{1}{*}{--}} & \multicolumn{1}{c|}{\multirow{1}{*}{--}} & \multicolumn{1}{l|}{$A_{50}$} & $\phi_{15}$ & $(J_1 - 3)^5$            \\
			                                           & \multicolumn{1}{c|}{\multirow{1}{*}{--}} & \multicolumn{1}{c|}{\multirow{1}{*}{--}} & \multicolumn{1}{c|}{\multirow{1}{*}{--}} & \multicolumn{1}{c|}{\multirow{1}{*}{--}} & \multicolumn{1}{c|}{\multirow{1}{*}{--}} & \multicolumn{1}{c|}{\multirow{1}{*}{--}}       & \multicolumn{1}{c|}{\multirow{1}{*}{--}} & \multicolumn{1}{c|}{\multirow{1}{*}{--}} & \multicolumn{1}{l|}{$A_{41}$} & $\phi_{16}$ & $(J_1 - 3)^4(J_2 - 3)$   \\
			                                           & \multicolumn{1}{c|}{\multirow{1}{*}{--}} & \multicolumn{1}{c|}{\multirow{1}{*}{--}} & \multicolumn{1}{c|}{\multirow{1}{*}{--}} & \multicolumn{1}{c|}{\multirow{1}{*}{--}} & \multicolumn{1}{c|}{\multirow{1}{*}{--}} & \multicolumn{1}{c|}{\multirow{1}{*}{--}}       & \multicolumn{1}{c|}{\multirow{1}{*}{--}} & \multicolumn{1}{c|}{\multirow{1}{*}{--}} & \multicolumn{1}{l|}{$A_{32}$} & $\phi_{17}$ & $(J_1 - 3)^3(J_2 - 3)^2$ \\
			                                           & \multicolumn{1}{c|}{\multirow{1}{*}{--}} & \multicolumn{1}{c|}{\multirow{1}{*}{--}} & \multicolumn{1}{c|}{\multirow{1}{*}{--}} & \multicolumn{1}{c|}{\multirow{1}{*}{--}} & \multicolumn{1}{c|}{\multirow{1}{*}{--}} & \multicolumn{1}{c|}{\multirow{1}{*}{--}}       & \multicolumn{1}{c|}{\multirow{1}{*}{--}} & \multicolumn{1}{c|}{\multirow{1}{*}{--}} & \multicolumn{1}{l|}{$A_{23}$} & $\phi_{18}$ & $(J_1 - 3)^2(J_2 - 3)^3$ \\
			                                           & \multicolumn{1}{c|}{\multirow{1}{*}{--}} & \multicolumn{1}{c|}{\multirow{1}{*}{--}} & \multicolumn{1}{c|}{\multirow{1}{*}{--}} & \multicolumn{1}{c|}{\multirow{1}{*}{--}} & \multicolumn{1}{c|}{\multirow{1}{*}{--}} & \multicolumn{1}{c|}{\multirow{1}{*}{--}}       & \multicolumn{1}{c|}{\multirow{1}{*}{--}} & \multicolumn{1}{c|}{\multirow{1}{*}{--}} & \multicolumn{1}{l|}{$A_{14}$} & $\phi_{19}$ & $(J_1 - 3)(J_2 - 3)^4$   \\
			                                           & \multicolumn{1}{c|}{\multirow{1}{*}{--}} & \multicolumn{1}{c|}{\multirow{1}{*}{--}} & \multicolumn{1}{c|}{\multirow{1}{*}{--}} & \multicolumn{1}{c|}{\multirow{1}{*}{--}} & \multicolumn{1}{c|}{\multirow{1}{*}{--}} & \multicolumn{1}{c|}{\multirow{1}{*}{--}}       & \multicolumn{1}{c|}{\multirow{1}{*}{--}} & \multicolumn{1}{c|}{\multirow{1}{*}{--}} & \multicolumn{1}{l|}{$A_{05}$} & $\phi_{20}$ & $(J_2 - 3)^5$            \\ \hline
			\multicolumn{1}{|l|}{$N_{\text{vol}} = 1$} & \multicolumn{1}{l|}{$B_1$}               & $\phi_{3}$                               & \multicolumn{1}{l|}{$B_1$}               & $\phi_{6}$                               & \multicolumn{1}{l|}{$B_1$}               & $\phi_{10}$                                    & \multicolumn{1}{l|}{$B_1$}               & $\phi_{15}$                              & \multicolumn{1}{l|}{$B_1$}    & $\phi_{21}$ & $(J_3 - 1)^2$            \\ \hline
			\multicolumn{1}{|l|}{$N_{\text{vol}} = 2$} & \multicolumn{1}{l|}{$B_2$}               & $\phi_{4}$                               & \multicolumn{1}{l|}{$B_2$}               & $\phi_{7}$                               & \multicolumn{1}{l|}{$B_2$}               & $\phi_{11}$                                    & \multicolumn{1}{l|}{$B_2$}               & $\phi_{16}$                              & \multicolumn{1}{l|}{$B_2$}    & $\phi_{22}$ & $(J_3 - 1)^4$            \\ \hline
		\end{tabular}
		\caption{The coefficients \( A_{ij} \) and \( B_i \) for the strain energy density function \( W \) in the case where \( N_{\text{MR}} \in \{ 1, 2, 3, 4, 5 \} \) and \( N_{\text{vol}} \in \{ 1, 2 \} \).}
		\label{tab:PaperC_coefficients}
	\end{table}

	\noindent The strain energy density function is assumed to consist of both isochoric and volumetric contributions. Given \( N_{\text{MR}} = 5 \) and \( N_{\text{vol}} = 2 \), the number of isochoric and volumetric basis functions, respectively, the complete set of basis functions is given in \autoref{tab:PaperC_coefficients}.
	In this contribution, we choose the case where \( N_{\text{MR}} = 3 \) and \( N_{\text{vol}} = 1 \). Therefore, the strain energy density function is defined by:
	\begin{equation}
		\begin{split}
			W =\; & A_{10}(J_1 - 3) + A_{01}(J_2 - 3) + A_{20}(J_1 - 3)^2 + A_{11}(J_1 - 3)(J_2 - 3) \\
			& + A_{02}(J_2 - 3)^2 + A_{30}(J_1 - 3)^3 + A_{21}(J_1 - 3)^2(J_2 - 3)             \\
			& + A_{12}(J_1 - 3)(J_2 - 3)^2 + A_{03}(J_2 - 3)^3 + B_1(J_3 - 1)^2
		\end{split}
		\label{eq:PaperC_W_MR_3}
	\end{equation}
	To embed the strain energy density function constructed from the chosen basis functions into the finite element formulation, we first need to compute the second Piola--Kirchhoff stress tensor. For hyperelastic materials, this stress tensor is derived from the strain energy density \( W \) as
	\begin{equation}
		\secondPK = \frac{\partial W}{\partial \tenE},
	\end{equation}
	where \( \tenE = \frac{1}{2} \left( \matC - \tenI \right) \) is the Lagrangian (Green) strain tensor, with \( \matC = \defGrad^\top \defGrad \) being the right Cauchy--Green deformation tensor, and \( \tenI \) the identity tensor. The second Piola--Kirchhoff stress tensor is then given by:
	\begin{equation}
		\begin{split}
			\mathbf{S} =\; & A_{10} \, \mathbf{J}_{1,\mathbf{E}} + A_{01}\, \mathbf{J}_{2,\mathbf{E}} + 2A_{20}\,(J_1 - 3)\, \mathbf{J}_{1,\mathbf{E}} + A_{11}\,(J_2 - 3) \, \mathbf{J}_{1,\mathbf{E}} \\
			& + A_{11}\,(J_1 - 3) \, \mathbf{J}_{2,\mathbf{E}} + 2A_{02}\,(J_2 - 3)\, \mathbf{J}_{2,\mathbf{E}} + 3A_{30}\,(J_1 - 3)^2 \,\mathbf{J}_{1,\mathbf{E}}                       \\
			& + A_{21}\,(J_1 - 3)^2 \,\mathbf{J}_{2,\mathbf{E}} + 2A_{21}\,(J_1 - 3)\,(J_2 - 3)\, \mathbf{J}_{1,\mathbf{E}}                                                              \\
			& + 2A_{12}\,(J_1 - 3)\,(\,J_2 - 3) \,\mathbf{J}_{2,\mathbf{E}} + A_{12}\,(J_2 - 3)^2 \,\mathbf{J}_{1,\mathbf{E}}                                                            \\
			& + 3A_{03}\,(J_2 - 3)^2 \,\mathbf{J}_{2,\mathbf{E}} + 2B_1\,(J_3 - 1) \, \mathbf{J}_{3,\mathbf{E}}
		\end{split}
		\label{eq:PaperC_S_MR_3}
	\end{equation}
	The consistent material (tangent) stiffness tensor \( \mathbb{D} \), required for the finite element implementation, is obtained by differentiating the second Piola--Kirchhoff stress tensor w.r.t. the Green-Lagrange strain tensor:
	\begin{equation}
		\begin{split}
			\mathbb{D} & = \frac{\partial \secondPK}{\partial \tenE}                                                                                                                                                                                                                                                          \\
			& =\; A_{10} \, \mathbf{J}_{1,\mathbf{EE}} + A_{01} \, \mathbf{J}_{2,\mathbf{EE}} + 2 A_{20}\, \mathbf{J}_{1,\mathbf{E}} \otimes \mathbf{J}_{1,\mathbf{E}} + 2 A_{20}\,(J_1 - 3)\, \mathbf{J}_{1,\mathbf{EE}} + A_{11}\, \mathbf{J}_{2,\mathbf{E}} \otimes \mathbf{J}_{1,\mathbf{E}}                   \\
			& + A_{11}\, \mathbf{J}_{1,\mathbf{E}} \otimes \mathbf{J}_{2,\mathbf{E}} + A_{11}\,(J_2 - 3)\, \mathbf{J}_{1,\mathbf{EE}} + A_{11}\,(J_1 - 3) \,\mathbf{J}_{2,\mathbf{EE}} + 2 A_{02}\; \mathbf{J}_{2,\mathbf{E}} \otimes \mathbf{J}_{2,\mathbf{E}} + 2 A_{02}\;(J_2 - 3)\; \mathbf{J}_{2,\mathbf{EE}} \\
			& + 6 A_{30}\;(J_1 - 3)\; \mathbf{J}_{1,\mathbf{E}} \otimes \mathbf{J}_{1,\mathbf{E}} + 3 A_{30}\;(J_1 - 3)^2\; \mathbf{J}_{1,\mathbf{EE}} + A_{21}\;(J_1 - 3)^2\; \mathbf{J}_{2,\mathbf{EE}} + 2 A_{21}\;(J_1 - 3)\;(J_2 - 3)\; \mathbf{J}_{1,\mathbf{EE}}                                            \\
			& + 2 A_{21}\;(J_1 - 3)\; \mathbf{J}_{1,\mathbf{E}} \otimes \mathbf{J}_{2,\mathbf{E}} + 2 A_{21}\;(J_2 - 3)\; \mathbf{J}_{1,\mathbf{E}} \otimes \mathbf{J}_{1,\mathbf{E}} + 2 A_{21}\;(J_1 - 3)\; \mathbf{J}_{2,\mathbf{E}} \otimes \mathbf{J}_{1,\mathbf{E}}                                          \\
			& + 2 A_{12}\;(J_2 - 3) \;\mathbf{J}_{1,\mathbf{E}} \otimes \mathbf{J}_{2,\mathbf{E}} + 2 A_{12}\;(J_1 - 3) \;\mathbf{J}_{2,\mathbf{E}} \otimes \mathbf{J}_{2,\mathbf{E}} + A_{12}\;(J_2 - 3)^2 \;\mathbf{J}_{1,\mathbf{EE}}                                                                           \\
			& + 2 A_{12}\;(J_1 - 3)\;(J_2 - 3) \;\mathbf{J}_{2,\mathbf{EE}} + 2 A_{12}\;(J_2 - 3) \;\mathbf{J}_{2,\mathbf{E}} \otimes \mathbf{J}_{1,\mathbf{E}} + 6 A_{03}\;(J_2 - 3) \;\mathbf{J}_{2,\mathbf{E}} \otimes \mathbf{J}_{2,\mathbf{E}}                                                                \\
			& + 3 A_{03}\;(J_2 - 3)^2 \;\mathbf{J}_{2,\mathbf{EE}} + 2 B_1\;(J_3 - 1) \, \mathbf{J}_{3,\mathbf{EE}} + 2 B_1 \, \mathbf{J}_{3,\mathbf{E}} \otimes \mathbf{J}_{3,\mathbf{E}}
		\end{split}
		\label{eq:PaperC_D_MR_3}
	\end{equation}
	Here, \( \mathbf{J}_{i,\mathbf{E}} \) and \( \mathbf{J}_{i,\mathbf{EE}} \) are the first and second derivatives of the invariants for \( \mathbf{E} \), respectively.

	\section{Maximum a posteriori estimation of state and material parameters}\label{sec:map}

	\newcommand{\rs}{\vecr_{\mathrm{s}}}
	\newcommand{\re}{\vecr_{\mathrm{e}}}

	The all-at-once formulation in~\eqref{eq:PaperC_coupled_eq} can be interpreted as a MAP estimate arising from a Bayesian treatment of the stochastic state and observation equations. To make this connection explicit, we begin by recalling that the governing equations can be expressed in residual form:
	\begin{align}
		\vecr_{\mathrm{s}}(\veckappa, \vecu, \vecX_\text{P}) & = \pmb{\mathcal{F}}(\vecu, \veckappa), \label{eq:error_state} \\[0.5ex]
		\vecr_{\mathrm{e}}(\vecu, \vecy)                     & = \Rev{\matH} \, \vecu-\vecy, \label{eq:error_obs}
	\end{align}
	where $\vecX_{\text{P}}\in\mathbb{R}^{\ngdof}$ denote the finite element nodal points, $\vecr_{\mathrm{s}}$ denotes the residual of the state equation and $\vecr_{\mathrm{e}}$ the residual of the observation equation. Note that in \eqref{eq:error_obs}, the noise $\vece$ is contained in the residual $\vecr_{\mathrm{e}}$.

	We are then interested in the posterior
	\begin{equation}\label{eq:posterior}
		f_{\vecu,\veckappa \mid \rs, \re}(\vecu,\veckappa \mid \rs, \re) =  \frac{f_{\rs, \re\mid \vecu, \veckappa}(\rs, \re \mid \vecu, \veckappa) f_{\vecu, \veckappa}(\vecu, \veckappa)}{f_{\rs, \re}(\rs, \re)},
	\end{equation}
	where $f_{\vecu, \veckappa}(\vecu, \veckappa)$ denotes the joint prior distribution of the state and parameters.
	The likelihood $f_{\rs, \re \mid \vecu, \veckappa}$ can be expanded as
	\begin{equation}\label{eq:likelihood_split}
		f_{\rs, \re \mid \vecu, \veckappa}(\rs, \re \mid \vecu, \veckappa) = f_{\rs \mid \vecu, \veckappa}(\rs \mid \vecu, \veckappa) \, f_{\re\mid \vecu}(\re\mid \vecu),
	\end{equation}
	and the posterior \eqref{eq:posterior} becomes:
	\begin{equation}\label{eq:posterior_2}
		f_{\vecu,\veckappa \mid \rs, \re}(\vecu,\veckappa \mid \rs, \re) \;\propto\; f_{\rs \mid \vecu, \veckappa}(\rs \mid \vecu, \veckappa) \, \, f_{\re\mid \vecu}(\re\mid \vecu)\,\, f_{\vecu, \veckappa}(\vecu, \veckappa)
	\end{equation}
	Assuming Gaussian error models for the residuals $\vecr_{\mathrm{s}}$ \eqref{eq:error_state} and $\vecr_{\mathrm{e}}$  \eqref{eq:error_obs}, we may write the likelihoods as
	\begin{equation}\label{eq:distribution}
		\begin{aligned}
			\vecr_{\mathrm{s}}(\veckappa, \vecu, \vecX_\text{P}) & \sim f_{\rs \mid \vecu,\veckappa}(\rs \mid \vecu,\veckappa) & = \mathcal{N}(\vec0, \sigma_{\mathrm{s}}^2 \, \matI),                                           \\
			\vecr_{\mathrm{e}}(\vecu, \vecy)                     & \sim f_{\re \mid \vecu}(\re \mid \vecu)                     & = \mathcal{N}(\vec0, \sigma_{\mathrm{e}}^2 \, \matI)                                          ,
		\end{aligned}
	\end{equation}
	with $\sigma_{\text{e}}$ and $\sigma_{\text{s}}$ standard deviations of the noise and the state residuals. With \eqref{eq:distribution}, the MAP estimate of \eqref{eq:posterior_2} --- which is equivalent to the minimum of the negative log-posterior --- becomes:
	\begin{equation}\label{eq:map_general}
		\{\vecu^*,\veckappa^*\}
		= \arg\min_{\vecu,\veckappa}
		\left[
		\|\pmb{\mathcal{F}}(\vecu, \veckappa)\|^{2}_2
		+ \|\Rev{\matH} \, \vecu-\vecy\|^{2}_2
		- \ln{f_{\vecu, \veckappa}(\vecu, \veckappa)}
		\right].
	\end{equation}
	If an uninformative prior is chosen, the last term in \eqref{eq:map_general} vanishes. In addition, if the noise is neglected as in the deterministic setting, we recover the coupled optimization problem introduced in~\eqref{eq:PaperC_coupled_eq} up to the weighting factors $\omega_s$ and $\omega_d$. This probabilistic interpretation highlights that the proposed coupled formulation is not arbitrary: it can be regarded as the MAP estimator under Gaussian error assumptions.

	\section{The Nearest Symmetric Positive Definite Matrix}\label{sec:PaperC_nearest_spd}

	\noindent \textbf{Definition (Nearest SPD operator):}
	For any square matrix $\mathbf{M}\in\mathbb{R}^{n\times n}$, we define the \emph{nearest symmetric positive definite (SPD) operator} as
	\begin{equation}\label{eq:PaperC_nearest_spd_operator}
		\operatorname{SPD}(\mathbf{M})
		:= \argmin_{\substack{\mathbf{X}=\mathbf{X}^\top\\ \mathbf{X}\succ 0}}
		\|\mathbf{X}-\mathbf{M}\|_F ,
	\end{equation}
	where $\|\cdot\|_F$ denotes the Frobenius norm, and the minimization is taken over the set of symmetric positive definite matrices. Here, the notation $\mathbf{X} \succ 0$ enforces positive definiteness and $\mathbf{X} = \mathbf{X}^\top$ enforces symmetry. Thus, $\operatorname{SPD}(\mathbf{M})$ returns the closest SPD matrix to $\mathbf{M}$ in the Frobenius sense.

	\vspace{0.5em}
	\noindent \textbf{Numerical realization (Higham's method):}
	A practical and robust way to evaluate $\operatorname{SPD}(\mathbf{M})$ is based on the symmetric polar decomposition of the symmetrized matrix~\cite{higham1988computing}. The following steps compute the optimizer of \eqref{eq:PaperC_nearest_spd_operator} up to machine precision:

	\begin{enumerate}
		\item \textbf{Symmetrize:} Form the symmetric part  $\mathbf{B} \;=\;\tfrac12\bigl(\mathbf{M}+\mathbf{M}^\top\bigr)$
		\item \textbf{Polar factor:} Compute an SVD of $\mathbf{B}$, $\mathbf{B}=\mathbf{U}\boldsymbol{\Sigma}\mathbf{V}^\top$, and $\mathbf{S} \;=\; \mathbf{V}\boldsymbol{\Sigma}\mathbf{V}^\top$ and then define the first projection iterate
		      \[
			      \mathbf{M}_0 \;=\; \tfrac12\bigl(\mathbf{B}+\mathbf{S}\bigr).
		      \]
		\item \textbf{Enforce strict positive definiteness:} If $\mathbf{M}_0$ is not strictly positive definite (e.g., the Cholesky factorization fails), add the smallest diagonal shift that achieves SPD:
		      \[
			      \mathbf{M}^\text{SPD} \;=\; \mathbf{M}_0 + \delta \mathbf{I},
			      \qquad \delta>0 \text{ (smallest shift ensuring positive definiteness)} .
		      \]
	\end{enumerate}

	\section{Additional Results: Strain Energy vs. Deformation Gradient Components}\label{sec:PaperC_additional_results}
\begin{figure}[!t]
		\centering
		\subfloat[$\nsen =13$, $\sigma_{\vece} = 10^{-03}$]{
			\centering
			\includegraphics{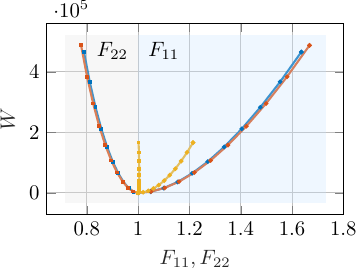}
			\hspace*{-0.4cm}
			\includegraphics{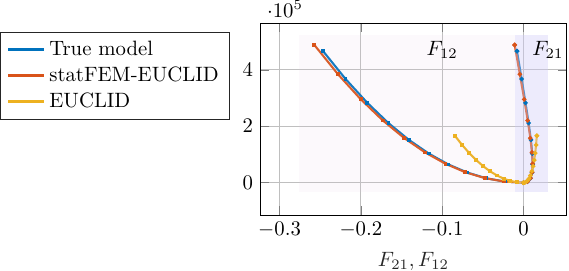}
			\label{subfig:plateWithHole_energy_vs_defGrad_nsen13_e3}} \\
		\subfloat[$\nsen =38$, $\sigma_{\vece} = 10^{-03}$]{
			\centering
			\includegraphics{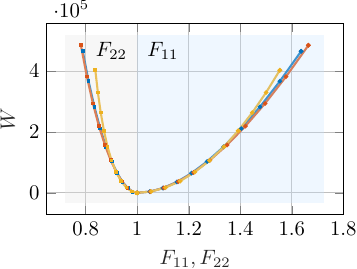}
			\hspace*{-0.4cm}
			\includegraphics{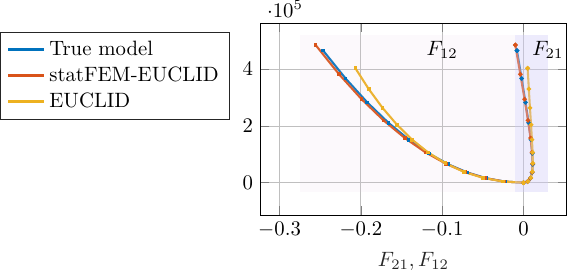}
			\label{subfig:plateWithHole_energy_vs_defGrad_nsen38_e3}}
		\caption{Comparison of the strain energy function $W$ between the true Neo--Hookean model, EUCLID, and statFEM-EUCLID for different sensor configurations and noise levels $\sigma_{\vece} = 10^{-3}$. Panel (a) shows results with $\nsen = 13$, while panel (b) corresponds to $\nsen = 38$. In all cases, statFEM-EUCLID provides a closer match to the ground truth than EUCLID.  The deformation gradient is controlled via the load scale $\eta\in[0,1]$, c.f.~\autoref{fig:plateWithHole_energy_vs_f}.  Results are depicted for the element marked in \autoref{fig:plateWithHole_mesh}.}
		\label{fig:plateWithHole_energy_vs_defGrad_e3}
	\end{figure}

\begin{figure}[!t]
		\centering
		\subfloat[$\nsen =13$, $\sigma_{\vece} = 10^{-04}$]{
			\centering
			\includegraphics{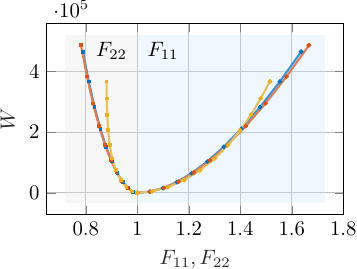}
			\hspace*{-0.4cm}
			\includegraphics{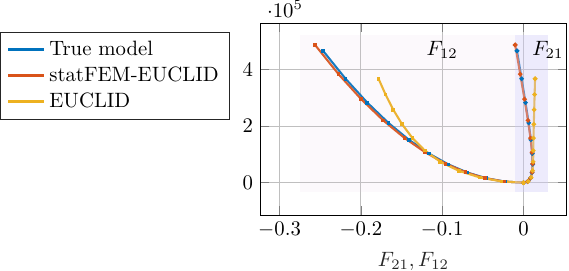}
			\label{subfig:plateWithHole_energy_vs_defGrad_nsen13_e4}} \\
		\subfloat[$\nsen =38$, $\sigma_{\vece} = 10^{-04}$]{
			\centering
			\includegraphics{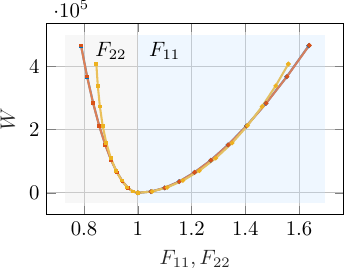}
			\hspace*{0cm}
			\includegraphics{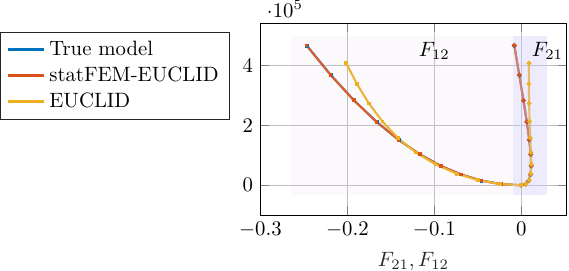}
			\label{subfig:plateWithHole_energy_vs_defGrad_nsen38_e4}}
		\caption{Comparison of the strain energy function $W$ between the true Neo--Hookean model, EUCLID, and statFEM-EUCLID for different sensor configurations and noise levels $\sigma_{\vece} = 10^{-4}$. Panel (a) shows results with $\nsen = 13$, while panel (b) corresponds to $\nsen = 38$. In all cases, statFEM-EUCLID provides a closer match to the ground truth than EUCLID, with the best agreement achieved for dense sensing ($n_{\text{sen}} = 38$) and low noise. The deformation gradient is controlled via the load scale $\eta\in[0,1]$, c.f.~\autoref{fig:plateWithHole_energy_vs_f}. Results are depicted for the element marked in \autoref{fig:plateWithHole_mesh}.}
		\label{fig:plateWithHole_energy_vs_defGrad_e4}
	\end{figure}

	The plots in~\autoref{fig:plateWithHole_energy_vs_defGrad_e3} and \autoref{fig:plateWithHole_energy_vs_defGrad_e4} compare the strain energy density function $W$ against the deformation gradient \( \matF \) on the representative element (see~\autoref{fig:plateWithHole_mesh}) predicted by the discovered models against the Neo--Hookean ground truth for two sensor configurations. Here, the deformation gradient is expressed as
	\[
		\matF =
		\begin{bmatrix}
			F_{11} & F_{12} \\
			F_{21} & F_{22}
		\end{bmatrix},
	\]
	where the diagonal components represent normal stretches and the off-diagonal components represent shear deformations. For $n_{\text{sen}} = 13$, the statFEM-EUCLID framework provides a much closer match to the true model than EUCLID, which exhibits noticeable deviation, particularly at larger stretches. When the number of sensors increases to $n_{\text{sen}} = 38$, both approaches improve, but statFEM-EUCLID continues to yield the most faithful recovery, with the discovered curve nearly overlapping the ground truth.

\end{appendices}
\bibliographystyle{unsrtnat}
\bibliography{literature}

@article{girolami2021statistical,
  title = {The Statistical Finite Element Method ({statFEM}) for Coherent
           Synthesis of Observation Data and Model Predictions},
  author = {Girolami, M. and Febrianto, E. and Yin, G. and Cirak, F.},
  journal = {Computer Methods in Applied Mechanics and Engineering},
  volume = {375},
  pages = {113533},
  year = {2021},
  publisher = {Elsevier},
}

@article{kennedy2001bayesian,
  title = {{Bayesian} Calibration of Computer Models},
  author = {Kennedy, M. C. and O'Hagan, A.},
  journal = {Journal of the Royal Statistical Society: Series B (Statistical
             Methodology)},
  volume = {63},
  number = {3},
  pages = {425--464},
  year = {2001},
  publisher = {Wiley Online Library},
}

@article{berveiller2006stochastic,
  title = {Stochastic Finite Element: a Non-Intrusive Approach by Regression},
  author = {Berveiller, M. and Sudret, B. and Lemaire, M.},
  journal = {European Journal of Computational Mechanics/Revue Europ{\'e}enne de
             M{\'e}canique Num{\'e}rique},
  volume = {15},
  number = {1-3},
  pages = {81--92},
  year = {2006},
  publisher = {Taylor \& Francis},
}

@inproceedings{berveiller2005non,
  title = {Non Linear Non Intrusive Stochastic Finite Element Method-Application
           to a Fracture Mechanics Problem},
  author = {Berveiller, M. and Sudret, B. and Lemaire, M.},
  booktitle = {Proc. 9th Int. Conf. Struct. Safety and Reliability
               (ICOSSAR'2005), Rome, Italie},
  year = {2005},
}

@article{reagana2003uncertainty,
  title = {Uncertainty Quantification in Reacting-flow Simulations through
           Non-intrusive Spectral Projection},
  author = {Reagana, M. T. and Najm, H. N. and Ghanem, R. G. and Knio, O. M.},
  journal = {Combustion and Flame},
  volume = {132},
  number = {3},
  pages = {545--555},
  year = {2003},
  publisher = {Elsevier},
}

@article{Arendt2012,
  author = {Arendt, P. D. and Apley, D. W. and Chen, W.},
  journal = {Journal of Mechanical Design, Transactions of the Asme},
  number = {10},
  pages = {1--12},
  title = {Quantification of Model Uncertainty: Calibration, Model Discrepancy,
           and Identifiability},
  Volume = {134},
  year = {2012},
}

@article{deb2001solution,
  title = {Solution of Stochastic Partial Differential Equations using Galerkin
           Finite Element Techniques},
  author = {Deb, M. K. and Babu{\v{s}}ka, I. M. and Oden, J. T.},
  journal = {Computer Methods in Applied Mechanics and Engineering},
  volume = {190},
  number = {48},
  pages = {6359--6372},
  year = {2001},
  publisher = {Elsevier},
}

@article{matthies2005galerkin,
  title = {Galerkin Methods for Linear and Nonlinear Elliptic Stochastic Partial
           Differential Equations},
  author = {Matthies, H. G. and Keese, A.},
  journal = {Computer Methods in Applied Mechanics and Engineering},
  volume = {194},
  number = {12-16},
  pages = {1295--1331},
  year = {2005},
  publisher = {Elsevier},
}

@article{koh2023stochastic,
  author = {Koh, K. J. and Cirak, F.},
  journal = {Computer Methods in Applied Mechanics and Engineering},
  pages = {116358},
  title = {Stochastic {PDE} Representation of Random Fields for Large-scale {
           Gaussian} Process Regression and Statistical Finite Element Analysis},
  volume = {417},
  year = {2023},
}

@article{akyildiz2022statistical,
  author = {Akyildiz, {\"O}. D. and Duffin, C. and Sabanis, S. and Girolami, M.},
  journal = {SIAM/ASA Journal on Uncertainty Quantification},
  pages = {1560--1585},
  title = {Statistical Finite Elements via {L}angevin Dynamics},
  volume = {10},
  year = {2022},
}

@article{narouie2023inferring,
  title = {Inferring Displacement Fields from Sparse Measurements Using the
           Statistical Finite Element Method},
  author = {Narouie, V. and Wessels, H. and R{\"o}mer, U.},
  journal = {Mechanical Systems and Signal Processing},
  volume = {200},
  pages = {110574},
  year = {2023},
  publisher = {Elsevier},
}

@article{grediac2006virtual,
  title = {The virtual fields method for extracting constitutive parameters from
           full-field measurements: a review},
  author = {Grediac, M. and Pierron, F. and Avril, S. and Toussaint, E.},
  journal = {Strain},
  volume = {42},
  number = {4},
  pages = {233--253},
  year = {2006},
  publisher = {Wiley Online Library},
}

@article{romer2025reduced,
  title = {Reduced and all-at-once approaches for model calibration and
           discovery in computational solid mechanics},
  author = {R{\"o}mer, U. and Hartmann, S. and Tr{\"o}ger, J.-A. and Anton, D.
            and Wessels, H. and Flaschel, M. and De Lorenzis, L.},
  journal = {Applied Mechanics Reviews},
  volume = {77},
  number = {4},
  pages = {040801},
  year = {2025},
  publisher = {American Society of Mechanical Engineers},
}

@article{narouie2025mechanical,
  title = {Mechanical state estimation with a Polynomial-Chaos-Based Statistical
           Finite Element Method},
  author = {Narouie, V. and Wessels, H. and Cirak, F. and R{\"o}mer, U.},
  journal = {Computer Methods in Applied Mechanics and Engineering},
  volume = {441},
  pages = {117970},
  year = {2025},
  publisher = {Elsevier},
}

@article{flaschel2021unsupervised,
  title = {Unsupervised discovery of interpretable hyperelastic constitutive
           laws},
  author = {Flaschel, M. and Kumar, S. and De Lorenzis, L.},
  journal = {Computer Methods in Applied Mechanics and Engineering},
  volume = {381},
  pages = {113852},
  year = {2021},
}

@book{pierron2012virtual,
  title = {The virtual fields method: extracting constitutive mechanical
           parameters from full-field deformation measurements},
  author = {Pierron, F. and Gr{\'e}diac, M.},
  year = {2012},
  publisher = {Springer Science \& Business Media},
}

@article{higham1988computing,
  title = {Computing a nearest symmetric positive semidefinite matrix},
  author = {Higham, J.},
  journal = {Linear Algebra and Its Applications},
  volume = {103},
  pages = {103--118},
  year = {1988},
  publisher = {Elsevier},
}

@article{lourencco2024indirect,
  title = {An indirect training approach for implicit constitutive modelling
           using recurrent neural networks and the virtual fields method},
  author = {Louren{\c{c}}o, R. and Georgieva, P. and Cueto, E. and
            Andrade-Campos, A.},
  journal = {Computer Methods in Applied Mechanics and Engineering},
  volume = {425},
  pages = {116961},
  year = {2024},
  publisher = {Elsevier},
}

@article{wang2021inference,
  title = {Inference of deformation mechanisms and constitutive response of soft
           material surrogates of biological tissue by full-field
           characterization and data-driven variational system identification},
  author = {Wang, Z. and Estrada, J.B. and Arruda, E.M. and Garikipati, K.},
  journal = {Journal of the Mechanics and Physics of Solids},
  volume = {153},
  pages = {104474},
  year = {2021},
  publisher = {Elsevier},
}

@article{joshi2022bayesian,
  title = {{Bayesian}-{EUCLID}: Discovering hyperelastic material laws with
           uncertainties},
  author = {Joshi, A. and Thakolkaran, P. and Zheng, Y. and Escande, M. and
            Flaschel, M. and De Lorenzis, L. and Kumar, S.},
  journal = {Computer Methods in Applied Mechanics and Engineering},
  volume = {398},
  pages = {115225},
  year = {2022},
  publisher = {Elsevier},
}

@article{kissas2024language,
  title = {The language of hyperelastic materials},
  author = {Kissas, G. and Mishra, S. and Chatzi, E. and De Lorenzis, L.},
  journal = {Computer Methods in Applied Mechanics and Engineering},
  volume = {428},
  pages = {117053},
  year = {2024},
  publisher = {Elsevier},
}

@article{fuhg2024review,
  title = {A review on data-driven constitutive laws for solids},
  author = { Fuhg, J.N. and Anantha Padmanabha, G. and Bouklas, N. and Bahmani,
            B. and Sun, W. and Vlassis, N.N. and Flaschel, M. and Carrara, P. and
            De Lorenzis, L.},
  year = {2025},
  journal = {Archives of Computational Methods in Engineering},
  volume = {32},
  pages = {1841-1883},
  publisher = {Springer},
}

@incollection{sutton2008DIC,
  title = {Digital image correlation for shape and deformation measurements},
  author = {Sutton, M.A.},
  booktitle = {Springer Handbook of Experimental Solid Mechanics},
  pages = {565--600},
  year = {2008},
  publisher = {Springer},
}

@article{pierron2023MatTesting2,
  title = {Material Testing 2.0: A brief review},
  author = {Pierron, F.},
  journal = {Strain},
  volume = {59},
  number = {3},
  pages = {e12434},
  year = {2023},
  publisher = {Wiley Online Library},
}

@article{li2022equilibriumCNN,
  title = {Equilibrium-based convolution neural networks for constitutive
           modeling of hyperelastic materials},
  author = { Li, L. and Chen, C.Q. },
  journal = {Journal of the Mechanics and Physics of Solids},
  volume = {164},
  pages = {104931},
  year = {2022},
  publisher = {Elsevier},
}

@article{he2024review,
  title = {Review of research progress and development trend of digital image
           correlation},
  author = {He, X. and Zhou, R. and Liu, Z. and Yang, S. and Chen, K. and Li, L.
            },
  journal = {Multidiscipline Modeling in Materials and Structures},
  volume = {20},
  number = {1},
  pages = {81--114},
  year = {2024},
  publisher = {Emerald Publishing Limited},
}

@article{bogusz2024application,
  title = {Application of digital image correlation for strain mapping of
           structural elements and materials},
  author = {Bogusz, P. and Kraso{\'n}, W. and Pazur, K.},
  journal = {Materials},
  volume = {17},
  number = {11},
  pages = {2577},
  year = {2024},
  publisher = {MDPI},
}

@article{tu2020stress,
  title = {Stress--strain curves of metallic materials and post-necking strain
           hardening characterization: A review},
  author = {Tu, S. and Ren, X. and He, J. and Zhang, Z.},
  journal = {Fatigue \& Fracture of Engineering Materials \& Structures},
  volume = {43},
  number = {1},
  pages = {3--19},
  year = {2020},
  publisher = {Wiley Online Library},
}

@article{flaschel2023automated,
  title = {Automated discovery of generalized standard material models with {
           EUCLID}},
  author = {Flaschel, M. and Kumar, S. and De Lorenzis, L.},
  journal = {Computer Methods in Applied Mechanics and Engineering},
  volume = {405},
  pages = {115867},
  year = {2023},
  publisher = {Elsevier},
}

@article{akerson2025learning,
  title = {Learning constitutive relations from experiments: 1. {PDE}
           constrained optimization},
  author = {Akerson, A. and Rajan, A. and Bhattacharya, K.},
  journal = {Journal of the Mechanics and Physics of Solids},
  pages = {106128},
  year = {2025},
  publisher = {Elsevier},
}

@article{flaschel2023automatedbrain,
  title = {Automated discovery of interpretable hyperelastic material models for
           human brain tissue with {EUCLID}},
  author = {Flaschel, M. and Yu, H. and Reiter, N. and Hinrichsen, J. and Budday
            , S. and Steinmann, P. and Kumar, S. and De Lorenzis, L.},
  journal = {Journal of the Mechanics and Physics of Solids},
  volume = {180},
  pages = {105404},
  year = {2023},
  publisher = {Elsevier},
}

@article{wang2021variational,
  title = {Variational system identification of the partial differential
           equations governing microstructure evolution in materials: Inference
           over sparse and spatially unrelated data},
  author = {Wang, Z. and Huan, X. and Garikipati, K.},
  journal = {Computer Methods in Applied Mechanics and Engineering},
  volume = {377},
  pages = {113706},
  year = {2021},
  publisher = {Elsevier},
}

@article{chen2024finite,
  title = {Finite element model updating for material model calibration: A
           review and guide to practice},
  author = {Chen, B. and Starman, B. and Halilovi{\v{c}}, M. and Berglund, L. A.
            and Coppieters, S.},
  journal = {Archives of Computational Methods in Engineering},
  volume={32},
  number={4},
  pages={2035--2112},
  year={2025},
  publisher={Springer}
}

@article{avril2004sensitivity,
  title = {Sensitivity of the virtual fields method to noisy data},
  author = {Avril, S. and Gr{\'e}diac, M. and Pierron, F.},
  journal = {Computational Mechanics},
  volume = {34},
  number = {6},
  pages = {439--452},
  year = {2004},
  publisher = {Springer},
}

@article{rossi2015effect,
  title = {Effect of {DIC} spatial resolution, noise and interpolation error on
           identification results with the {VFM}},
  author = {Rossi, M. and Lava, P. and Pierron, F. and Debruyne, D. and Sasso,
            M.},
  journal = {Strain},
  volume = {51},
  number = {3},
  pages = {206--222},
  year = {2015},
  publisher = {Wiley Online Library},
}

@article{hirsh2022sparsifying,
  title = {Sparsifying priors for {Bayesian} uncertainty quantification in model
           discovery},
  author = {Hirsh, S. M. and Barajas-Solano, D. A. and Kutz, J. N.},
  journal = {Royal Society Open Science},
  volume = {9},
  number = {2},
  pages = {211823},
  year = {2022},
  publisher = {The Royal Society},
}

@article{fasel2022ensemble,
  title = {Ensemble-{SINDy}: Robust sparse model discovery in the low-data,
           high-noise limit, with active learning and control},
  author = {Fasel, U. and Kutz, J. N. and Brunton, B. W. and Brunton, S. L.},
  journal = {Proceedings of the Royal Society A},
  volume = {478},
  number = {2260},
  pages = {20210904},
  year = {2022},
  publisher = {The Royal Society},
}

@article{jiang2015surrogate,
  title = {Surrogate preposterior analyses for predicting and enhancing
           identifiability in model calibration},
  author = {Jiang, Z. and Apley, D. W. and Chen, W.},
  journal = {International Journal for Uncertainty Quantification},
  volume = {5},
  number = {4},
  pages = {341-359},
  year = {2015},
  publisher = {Begel House Inc.},
}

@article{thakolkaran2022nn,
  title = {{NN-EUCLID}: Deep-learning hyperelasticity without stress data},
  author = {Thakolkaran, P. and Joshi, A. and Zheng, Y. and Flaschel, M. and De
            Lorenzis, L. and Kumar, S.},
  journal = {Journal of the Mechanics and Physics of Solids},
  volume = {169},
  pages = {105076},
  year = {2022},
  publisher = {Elsevier},
}

@article{karvonen2025error,
  title = {Error analysis for a statistical finite element method},
  author = {Karvonen, T. and Cirak, F. and Girolami, M.},
  journal = {Journal of Multivariate Analysis},
  pages = {105468},
  year = {2025},
  publisher = {Elsevier},
}

@book{kim2014introduction,
  title = {Introduction to nonlinear finite element analysis},
  author = {Kim, N.-H.},
  year = {2014},
  publisher = {Springer Science \& Business Media},
}

@article{ricker_systematicFittingHyperelasticity_2023,
  title = {Systematic Fitting and Comparison of Hyperelastic Continuum Models
           for Elastomers},
  author = { Ricker, A. and Wriggers, P. },
  year = {2023},
  journal = {Archives of Computational Methods in Engineering},
  volume = {30},
  number = {3},
  pages = {2257-2288},
}

@article{urre2025MDisc,
  title = {Automated Constitutive Model Discovery by Pairing Sparse Regression
           Algorithms with Model Selection Criteria},
  author = {Urrea-Quintero, J.-H. and Anton, D. and Lorenzis, L. D. and Wessels,
            H.},
  journal={Computer Methods in Applied Mechanics and Engineering},
  volume={449},
  pages={118551},
  year={2026},
  publisher={Elsevier}
}

@article{mcculloch2024LPRegularizationModelDiscovery,
  title = {On sparse regression, {L}p-regularization, and automated model
           discovery},
  author = {McCulloch, J. A. and St. Pierre, S. R. and Linka, K. and Kuhl, E.},
  year = {2024},
  journal = {International Journal for Numerical Methods in Engineering},
  volume = {125},
  number = {14},
  pages = {e7481},
  doi = {https://doi.org/10.1002/nme.7481},
}

@book{brunton2022data,
  title = {Data-driven science and engineering: Machine learning, dynamical
           systems, and control},
  author = {Brunton, S. L. and Kutz, J N.},
  year = {2022},
  publisher = {Cambridge University Press},
}

@misc{knauf_narouie_2025_17249520,
  author       = {Knauf Narouie, V.},
  title        = {Code for the publication ``{U}nsupervised Constitutive Model Discovery from Sparse and Noisy Data''},
  howpublished = {Github: \url{https://github.com/VhI3/statFEM-EUCLID}},
  year         = 2025,
  publisher    = {Zenodo},
  note = {DOI: \url{https://doi.org/10.5281/zenodo.18300661}},
}

@article{abbasi2025discovery,
  title={Discovery of Hyperelastic Constitutive Laws from Experimental Data with EUCLID},
  author={Abbasi, A. and Ricci, M. and Carrara, P. and Flaschel, M. and Kumar, S. and Marfia, S. and De Lorenzis, L.},
  journal={arXiv preprint arXiv:2510.24747},
  year={2025}
}

@article{flaschel2025material,
  title={Material Fingerprinting: A shortcut to material model discovery without solving optimization problems},
  author={Flaschel, M. and Martonov{\'a}, D. and Veil, C. and Kuhl, E.},
  journal={arXiv preprint arXiv:2508.07831},
  year={2025}
}

@article{hartmann_generalizedStrainEnergyFunctions_2003,
    title     = {Polyconvexity of generalized polynomial-type hyperelastic strain energy functions for near-incompressibility},    
    author    = {
        Hartmann, S. and 
        Neff, P.
    },
    year      = {2003},
    journal   = {International Journal of Solids and Structures},
    volume    = {40},
    number    = {11},
    pages     = {2767-2791}
}

@article{seidl2022calibration,
  title={Calibration of elastoplastic constitutive model parameters from full-field data with automatic differentiation-based sensitivities},
  author={Seidl, D. T. and Granzow, B. N.},
  journal={International Journal for Numerical Methods in Engineering},
  volume={123},
  number={1},
  pages={69--100},
  year={2022},
  publisher={Wiley Online Library}
}
\end{document}